\documentclass[11pt,a4paper]{article}
\usepackage{tikz}
\usepackage{wrapfig}
\usepackage{jheppub}
\usepackage{bbm}
\usepackage{youngtab}
\usepackage{rotfloat}
\usepackage{stmaryrd}
\usepackage{amsfonts,amssymb,amsmath}
\usepackage{mathrsfs}
\usepackage{hyperref}
\usepackage{tikz-cd}
\usepackage{tcolorbox}
\usepackage{lscape}

\def\a{\alpha}
\def\b{\beta}

\def\e{\epsilon}

\newcommand{\bC}{\ensuremath{\mathbb{C}}}

\newcommand{\bP}{\ensuremath{\mathbb{P}}}

\newcommand{\bR}{\ensuremath{\mathbb{R}}}

\newcommand{\bZ}{\ensuremath{\mathbb{Z}}}

\newcommand{\scP}{\ensuremath{\mathscr{P}}}

\newcommand{\frakS}{\ensuremath{\mathfrak{S}}}

\newcommand{\frakX}{\ensuremath{\mathfrak{X}}}

\newcommand{\fraksl}{\ensuremath{\mathfrak{sl}}}

\newcommand{\cF}{\mathcal{F}}

\newcommand{\cH}{\mathcal{H}}

\newcommand{\ccL}{{\mathcal{L}}}
\newcommand{\cM}{\mathcal{M}}
\newcommand{\cN}{\mathcal{N}}
\newcommand{\cO}{\mathcal{O}}

\newcommand{\cV}{\mathcal{V}}
\newcommand{\cX}{\mathcal{X}}

\newcommand{\cW}{\mathcal{W}}
\newcommand{\cR}{\mathcal{R}}

\newcommand{\SU}{\mathrm{SU}}
\newcommand{\SO}{\mathrm{SO}}

\newcommand{\Sp}{\mathrm{Sp}}
\newcommand{\U}{\mathrm{U}}

\newcommand{\Tr}{\mbox{Tr}}

\newcommand{\wh}{\widehat}

\newcommand{\dd}{\textrm{d}}

\newcommand{\rCS}{\textrm{rCS}}
\newcommand{\defo}{\textrm{def}}

\newcommand{\Ind}{\textrm{Ind}}

\def\bee{\begin{eqnarray*}}
\def\eee{\end{eqnarray*}}
\def\beee{\begin{equation*}}
\def\eeee{\end{equation*}}
\def\bea{\begin{eqnarray}}
\def\eea{\end{eqnarray}}
\def\be{\begin{equation}}
\def\ee{\end{equation}}
\def\ba{\begin{align}}
\def\ea{\end{align}}

\newcommand{\unknot}{{\raisebox{-.11cm}{\includegraphics[width=.37cm]{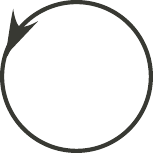}}}\,}

\preprint{CALT-2017-010}

\title{Refined large N duality for knots}

\author[1]{Masaya Kameyama}
\author[2,3,4]{and Satoshi Nawata}

\affiliation[1]{Graduate School of Mathematics, Nagoya University, Nagoya, 464-8602, Japan}
\affiliation[2]{Department of Physics and Center for Field Theory and Particle Physics, Fudan University,
220 Handan Road, 200433 Shanghai, China}
\affiliation[3]{Walter Burke Institute for Theoretical Physics, California Institute of Technology, Pasadena, CA 91125, USA}
\affiliation[4]{Centre for Quantum Geometry of Moduli Spaces, University of Aarhus, DK-8000, Denmark}

\emailAdd{m13020v@math.nagoya-u.ac.jp}
\emailAdd{snawata@gmail.com}

\abstract{
We formulate large $N$ duality of $\U(N)$ refined Chern-Simons theory with a torus knot/link in $S^3$. By studying refined BPS states in M-theory, we provide the explicit form of low-energy effective actions of Type IIA string theory with D4-branes on the $\Omega$-background. This form enables us to relate refined Chern-Simons invariants of a torus knot/link in $S^3$ to refined BPS invariants in the resolved conifold. Assuming that the extra $\U(1)$ global symmetry acts on BPS states trivially, the duality predicts graded dimensions of cohomology groups of moduli spaces of M2-M5 bound states associated to a torus knot/link in the resolved conifold. Thus, this formulation can be interpreted as a positivity conjecture of refined Chern-Simons invariants of torus knots/links. We also discuss about an extension to non-torus knots.
}

\dedicated{Dedicated to John H. Schwarz on his 75th birthday}

\begin{document}
\Yboxdim4pt
\maketitle

\section{Introduction}

Large $N$ duality is an equivalence between a $\U(N)$ gauge theory at large $N$ and a string theory. The original idea was due to 't Hooft \cite{tHooft:1973alw} and we have witnessed various successful incarnations of the idea in string theory. Among them, Gopakumar and Vafa proposed in the celebrated paper \cite{Gopakumar:1998ki} that the large $N$ limit of $\U(N)$ Chern-Simons theory on $S^3$ is equivalent to topological string theory on the resolved conifold. Since Chern-Simons theory is realized on  topological branes wrapped on $S^3$ in the deformed conifold $T^*S^3$ \cite{Witten:1992fb}, this duality can be also interpreted as \emph{geometric transition} at large $N$ in string theory. This proposal has far-reaching consequences both in physics and in mathematics. One of significant  consequences is a striking connection between two seemingly different theories of invariants. On the one hand, Chern-Simons theory provides a natural framework of quantum invariants of three-manifolds and knots \cite{Witten:1988hf}. On the other hand, topological string theory on a Calabi-Yau three-fold is mathematically formulated as theories of enumerative invariants involving some moduli spaces of curves in the three-fold. In particular, Labastida, Mari\~no, Ooguri and Vafa (LMOV)  \cite{Ooguri:1999bv,Labastida:2000zp,Labastida:2000yw,Labastida:2001ts} have put forth a remarkable relationship between quantum knot invariants and enumerative integral invariants in the resolved conifold by incorporating a knot in the duality.

As a parallel development, a theory of knot homology, which categorifies quantum invariants, has been constructed by Khovanov and Rozansky \cite{Khovanov:2000,Khovanov:2004,Khovanov:2005}, which enriches the large $N$ duality to a great extent. In fact, Gukov, Schwarz and Vafa \cite{Gukov:2004hz} proposed that $\fraksl(N)$ knot homology \cite{Khovanov:2004} at large $N$ can be identified with the space of BPS states in the resolved conifold, which can be understood as triply-graded HOMFLY-PT homology \cite{Khovanov:2005}. Furthermore, Aganagic and Shakirov have formulated refined Chern-Simons theory by using $\U(1)$ symmetry of a Seifert manifold or a torus knot \cite{Aganagic:2011sg,Aganagic:2012ne}. Remarkably, the large $N$ limits of refined Chern-Simons invariants of a torus knot  are conjectured to be equal to Poincar\'e polynomials of the corresponding HOMFLY-PT homology when colors are labelled by rectangular Young diagrams. In addition, refined Chern-Simons theory on $S^3$ at large $N$ is dual to refined topological string theory on the resolved conifold. Based on these results, refined topological vertex formulations are reproduced by using refined Chern-Simons theory at large $N$ \cite{Aganagic:2012hs}.

The goal of this paper is to formulate large $N$ duality of refined Chern-Simons theory with a torus knot. In other words, we shall find a relation between refined Chern-Simons invariants of a torus knot and refined BPS states in the resolved conifold, which is the generalization of the work of LMOV in refined topological string theory. This formulation also extends the proposal of \cite{Gukov:2004hz} to any colors in the case of torus knots. Since the form of partition functions on the deformed conifold has been determined in \cite{Aganagic:2011sg,Aganagic:2012ne,Aganagic:2012hs}, we shall investigate the form of partition functions on the resolved conifold and will find implications of the large $N$ duality on refined BPS spectra.

\vspace{.5cm}

To this end, let us first recall relevant setups of M-theory in which refined topological string theories are embedded.
On the deformed conifold side, we will consider the following brane configurations:

\vspace{-.2cm}
\noindent
\begin{small}
\begin{minipage}[b]{7.5cm}
\be
\begin{matrix}
{\mbox{\rm space-time:}} & \quad & S^1 & \times & TN_4 & \times  &T^* S^3 \\
{\mbox{\rm $N$ M5-branes:}} & \quad & S^1 &\times& D_q & \times &  S^3  \\
{\mbox{\rm $M$  M5'-branes:}} & \quad&  S^1 &\times& D_q & \times & \mathcal{N}_K
\end{matrix}\nonumber
\ee
\end{minipage}
\begin{minipage}[b]{7.5cm}
\be\label{deformed}
\begin{matrix}
{\mbox{\rm space-time:}} & \quad & S^1 & \times & TN_4 & \times  &T^* S^3 \\
{\mbox{\rm $N$ M5-branes:}} & \quad & S^1 &\times& D_q & \times &  S^3  \\
{\mbox{\rm $M$ M5'-branes:}} & \quad&  S^1 &\times& D_{\bar t} & \times & \mathcal{N}_K
\end{matrix}.
\ee
\end{minipage}
\end{small}
\vspace{.2cm}

\noindent
Here, $D_q$ is the two-dimensional cigar and $D_{\bar t}$ is the two-dimensional base of the Taub-NUT space $TN_4$. Moreover, writing the local complex coordinate $z_1$ for $D_q$ and $z_2$ for $D_{\bar t}$, we turn on the $\Omega$-background by the action
\be\label{omega}
(z_1, z_2)\to (qz_1, t^{-1}z_2)~.
\ee
The $N$ M5-branes wrap the zero section (a special Lagrangian submanifold) of the cotangent bundle $T^*S^3$ of a three-sphere, realizing Chern-Simons theory with $\U(N)$ gauge group on $S^3$ \cite{Witten:1992fb}.  The $M$ probe M5'-branes are located at another Lagrangian submanifold, which is the co-normal bundle $\mathcal{N}_K  \subset T^*S^3$ to a knot $K \subset S^3$ where the knot $K$ is realized as the intersection of the two stacks of the M5-branes $K = S^3 \cap \mathcal{N}_K $ \cite{Ooguri:1999bv}. As in \cite{Aganagic:2012hs}, we consider two distinct probe M5'-branes; one spanning on $D_q$ and the other on $D_{\bar t}$. Although both the configurations preserve four supercharges,
the $q$ and ${\bar t}$ branes become topological branes and anti-branes, respectively, in topological string theory.

The groundbreaking proposal in \cite{Gopakumar:1998ki} is that, at large $N$, the geometry undergoes the transition where $S^3$ shrinks and $S^2=\bC\bP^1$ is blown up.  As $S^3$ shrinks, in order for the probe M5'-branes to avoid the singularity, the conormal bundle $\cN_K $ is lifted to the fiber direction and it no longer touches  $S^3$. We refer the reader to \cite{Diaconescu:2011xr} for detailed treatment. As a result, the Calabi-Yau three-fold becomes the resolved conifold $X=\cO(-1)\oplus \cO(-1)\to \bC\bP^1$ where the $N$ M5-branes turn into B-field flux supporting $\bC\bP^1$ and the $M$ M5'-branes are situated on a Lagrangian submanifold $\mathcal{L}_K \subset X$ keeping the information of the knot $K$:

\vspace{-.2cm}
\noindent
\begin{minipage}[b]{7.5cm}
\be
\begin{matrix}
{\mbox{\rm space-time:}} & \quad & S^1 & \times & TN_4 & \times & X \\
{\mbox{\rm $M$  M5'-branes:}} & \quad & S^1 & \times & D_q & \times & \mathcal{L}_K
\end{matrix}\nonumber
\ee
\end{minipage}
\begin{minipage}[b]{7.5cm}
\be\label{resolved}
\begin{matrix}
{\mbox{\rm space-time:}} & \quad & S^1 & \times & TN_4 & \times & X \\
{\mbox{\rm $M$  M5'-branes:}} & \quad & S^1 & \times & D_{\bar t} & \times & \mathcal{L}_K  \\
\end{matrix}.
\ee\end{minipage}
\vspace{.2cm}

\noindent Indeed the large $N$ duality can be summarized as an equivalence of partition functions. On the deformed conifold, the partition function is a generating function of refined Chern-Simons invariants at large $N$. On the other hand, the form of a partition function on the resolved conifold can be determined by adapting Schwinger calculation and it is expressed as a sum of contributions of all the BPS states arising from M2-branes attached to the M5'-branes \cite{Gopakumar:1998jq,Gopakumar:1998ii,Ooguri:1999bv,Labastida:2000yw}. Hence, in this paper, we will establish an explicit connection of refined Chern-Simons theory to M2-M5' bound BPS states. Moreover, in the refined context,  we will conjecture a positivity property for refined Chern-Simons invariants.

\vspace{.4cm}

The organization of this paper is as follows. In \S\ref{sec:deformed}, we review refined Chern-Simons theory and generating functions of refined Chern-Simons invariants of torus knots. In \S \ref{sec:resolved}, we shall perform Schwinger's computation for BPS states in M-theory to determine low-energy effective actions of Type IIA string theory in the refined context. We first consider M-theory without M5'-branes for illustrative purpose and then proceed to the case with M5'-branes. In \S \ref{sec:knots}, the large $N$ duality is proposed for torus knots at the refined level. This relates refined Chern-Simons invariants to graded dimensions of BPS states in the resolved conifold, which results in a positivity conjecture for refined Chern-Simons invariants. The positivity property attributes to trivial action of the extra $\U(1)$ symmetry on the BPS states. In \S\ref{sec:links}, the large $N$ duality is generalized to multi-component torus links. In \S\ref{sec:non-torus}, we see that the refined large $N$ duality can be extended to a certain class of non-torus knots. In \S\ref{sec:discussion}, we discuss related open problems.  In Appendix \ref{sec:app-a}, we present a brief summary of symmetric functions used in this paper. Appendix \ref{sec:app-b} provides explicit formulas for reformulated invariants in terms of refined Chern-Simons invariants by using the large $N$ duality. Appendix \ref{sec:app-c} contains tables for BPS degeneracies associated to torus knots/links in the resolved conifold.

\section{Deformed conifold side}\label{sec:deformed}
On both the deformed conifold and resolved conifold side, low-energy effective actions are determined by taking account of small fluctuations around trajectory of a BPS particle arising from an M2-brane.
On the deformed conifold side,  partition functions of refined topological string theory have been analyzed in \cite{Aganagic:2011sg,Aganagic:2012ne,Aganagic:2012hs}. In this section, we briefly review generating functions of refined Chern-Simons invariants and we refer the reader to the original papers \cite{Aganagic:2011sg,Aganagic:2012ne,Aganagic:2012hs} for more details.

The work of Witten \cite{Witten:1992fb} provides the way to interpret Chern-Simons partition functions as an index counting BPS states in M-theory. From the viewpoint of M-theory, BPS states arise from M2-branes bridging between the M5-branes and the M5'-branes, and
they propagate only along the M5-branes. In low energy, one can suppress $S^3 \subset T^*S^3$ so that these states propagate along $\bR\times D_q \subset \bR\times TN_4$ when $S^1$ is large. Thus, Chern-Simons partition functions can be identified with an index $\Tr (-1)^F q^{J_1-J_2}$ of the three-dimensional theory on $S^1\times D_q$ with  $\cN = 2$ supersymmetry where $J_1$ is the Cartan generator of the $\SO(3)$ Lorentz symmetry of $S^1\times D_q$ and $J_2$ is the generator of the $\U(1)_2$ $\cR$-symmetry in $\cN = 2$ supersymmetry. In fact, $J_1$ and $J_2$ can be considered as generators of the rotations around the $z_1$ and $z_2$ plane, respectively.

In \cite{Aganagic:2011sg}, it is argued that the extra $\U(1)_R$ symmetry exists when $K$ is a torus knot $T_{m,n}$ (more generally Seifert three-manifolds). Then, using the charge $S_R$ of this $\U(1)_R$ symmetry, a refined Chern-Simons invariant of a torus knot is defined by a refined index \footnote{Charges of $J_1$, $J_2$ and $S_R$ are normalized to be half-integers.}
$$
\overline{\rCS}_{\SU(N),\lambda}(T_{m,n}) \; ``="   \; {\rm Tr}\;{}_{\cH_\lambda(T_{m,n} )}\, (-1)^{F}\, q^{J_1-S_R}\, t^{S_R-{J}_2}~.
$$
where $``=" $ means the equality up to a suitable normalization. Hence, refined Chern-Simons theory is defined via the 3d/3d correspondence. In \S\ref{sec:open}, we shall provide more explanation about the refined index. These invariants can be computed by using the modular $S$ and $T$ matrices associated to Macdonald functions \cite{cherednik1995macdonald,kirillov1996inner,etingof1996representation} or double affine Hecke algebra \cite{Cherednik:2011nr}. Moreover, it is proven in \cite{Gorsky:2013jna} that there exists a unique rational function $\overline{\rCS}_{\lambda}(T_{m,n};a,q,t)$ which is the stable large $N$ limit of refined Chern-Simons invariants with $\SU(N)$ gauge group \footnote{The brane setting gives rise to $\U(N)$ gauge group instead of $\SU(N)$ gauge group. However, the $\U(1)$ part merely provides the correction due to the framing number as well as the linking number of a knot/link, which  play no role in this paper. Therefore, the difference between $\U(N)$ and $\SU(N)$ invariants will be ignored in this paper.} of a torus knot in the following sense:
$$
\overline{\rCS}_{\lambda}(T_{m,n};a=t^N,q,t) =\overline{\rCS}_{\SU(N),\lambda}(T_{m,n};q,t)~.
$$
The stable limit $\overline{\rCS}_{\lambda}(T_{m,n};a,q,t)$ is also called \emph{DAHA-superpolynomial} \cite{Cherednik:2011nr}.
For instance, the refined Chern-Simons invariants of the unknot
\be\label{unknot}
\overline{\rCS}_{\lambda}(\unknot;a,q,t) =\prod_{x\in\lambda}\frac{t^{l'(x)}-aq^{a'(x)}}{1-q^{a(x)}t^{l(x)+1}}
\ee
so that, at the $a=t^N$ specialization, it becomes the principal specialization of the Macdonald function
$$
\overline{\rCS}_{\lambda}(\unknot;a=t^N,q,t) =P_\lambda(t^\rho;q,t)=\overline{\rCS}_{\SU(N),\lambda}(\unknot;q,t)~,
$$
where $\rho$ is the Weyl vector of $\fraksl(N)$. (See Figure \ref{fig:YT} for the notation.)
When the color $\lambda$ is labelled by a rectangular Young diagram, it is conjectured that the invariant $\overline{\rCS}_{\lambda}(T_{m,n};a,q,t)$ with change of variables \eqref{cov} coincides with the Poincar\'e polynomial of the corresponding HOMFLY-PT homology.
It turns out that refined Chern-Simons invariants have surprisingly rich properties. Especially, it is proven in \cite{cherednik2016daha} that the reduced invariants\footnote{We denote the \emph{unreduced} invariants by $\overline{\rCS}_{\lambda}(T_{m,n};a,q,t)$ and the \emph{reduced} invariants by ${\rCS}_{\lambda}(T_{m,n};a,q,t)$ where they are related by
$$
\overline{\rCS}_{\lambda}(T_{m,n};a,q,t)= \overline{\rCS}_{\lambda}(\unknot;a,q,t) ~{\rCS}_{\lambda}(T_{m,n};a,q,t)~.
$$} satisfy the following properties:
\begin{itemize}
\item  mirror/transposition symmetry  \be\label{mirror-sym} \rCS_{\lambda^T}(T_{m,n};a,q,t)=\rCS_{\lambda}(T_{m,n};a,t^{-1},q^{-1})~.\ee

\item  refined exponential growth property
\bea\label{EGP}
 \rCS_{\sum_{i=1}^\ell \lambda_i \omega_i}(T_{m,n};a,q=1,t)& =& \prod_{i=1}^\ell\Big[\rCS_{ \omega_i}(T_{m,n};a,q=1,t)\Big]^{\lambda_i}~,
 \eea
 where $\omega_i$ are the fundamental weights of $\fraksl(N)$.
 \end{itemize}

Indeed, a partition function of refined topological string theory on the deformed conifold can be determined by taking account of BPS particles arising from annulus M2-branes stretched between $S^3$ and $\cN_{T_{m,n}}$.
For the $q$-brane setting in \eqref{deformed}, the M2-brane stretching between $S^3$ and $\cN_{T_{m,n}}$ gives rise to two bifundamental $\cN=2$ chiral multiplets, $\Phi$ and $\widetilde \Phi$, on $S^1\times D_q$ charged as $(\textbf{N},\overline{\textbf{M}})$ and  $(\overline{\textbf{N}},{\textbf{M}})$, respectively, under $\U(N)\times \U(M)$. The chiral multiplet $\Phi$ has charges $(0,0;-\frac 12)$ under $(J_1,J_2;S_R)$ and $\widetilde \Phi$ is neutral. The three-dimensional $\cN=2$ index can be evaluated by counting ``single-letter index'' \cite{Imamura:2011su,Kapustin:2011jm} where the chiral multiplet $\Phi$ contributes to the single-letter index by $(q^{1/2}t^{-1/2}-1)/(1-q)$ and the other $\widetilde \Phi$ yields  $(1-q^{1/2}t^{1/2})/(1-q)$. From the viewpoint of the three-dimensional theory on $S^1\times D_q$, the $M$ M5'-branes give rise to $\U(M)$ flavor symmetry, and a torus knot $T_{m,n}$ in $S^3$ contributes to the single-letter index via the 3d/3d correspondence. Therefore, putting the single-letter index invariant under $\U(N)\times \U(M)$ into the plethystic exponent, the refined index takes the form
$$
Z_{\defo,\SU(N)}^{q}=\exp \left(\sum_{d>0}\sum_\mu\frac{1}{d}\frac{t^{\frac d2}-t^{-\frac d2}}{q^{\frac d2}-q^{-\frac d2}} \;  \Ind_{T_{m,n},\mu}(q^d,t^d) \; s_\mu(x^{d})\right)~,
$$
where $\Ind_{T_{m,n},\mu}(q,t)$ is a contribution of the knot $T_{m,n}$ to the single-particle index charged under the flavor representation $\mu$. This expression can be expanded by the basis of holonomies of the $\U(M)$ group which are the Macdonald functions $P_\lambda(x;q,t)$ in refined Chern-Simons theory
\begin{equation}\label{def-q}
Z_{\defo,\SU(N)}^{ q}=\sum_\lambda   \; g_\lambda(q,t) \; \overline \rCS_{\SU(N),\lambda}(T_{m,n}; q,t) \;P_\lambda(x; q,t)~,
\end{equation}
where  the function $g_\lambda$ can be determined by using the unknot invariants and it turns out to be the Macdonald norm  defined in Appendix \ref{sec:app-a}.
At the unrefined limit $q=t$,  it reduces to the generating function of $\SU(N)$ quantum invariants $ \overline J_{\SU(N),\lambda}(T_{m,n}; q) $ first obtained in \cite{Ooguri:1999bv}
$$
Z_{\defo,\SU(N)}=\sum_\lambda  \overline J_{\SU(N),\lambda}(T_{m,n}; q) \; s_{\lambda}(x) ~,
$$
where $s_\lambda(x)$ are the Schur functions.

On the other hand, the $\bar t$-branes intersect with the $q$-branes at a point in $D_q$ in \eqref{deformed} so that annulus M2-branes stretched between $S^3$ and $\cN_{T_{m,n}}$ bring about only a single bifundamental fermionic particle. Since it contributes to the refined index by $-1$, the index is of the form
$$
Z_{\defo,\SU(N)}^{\bar t}=\exp \left(\sum_{d>0}\sum_{\mu}\frac{(-1)}{d}\; \Ind_{T_{m,n},\mu}(q^d,t^d) \; s_{\mu}(x^{d})\right)~.
$$
For the $\bar t$-brane setting, the expansion in terms of the basis of holonomies  of the $\U(M)$ group  is more subtle. Since the $\bar t$-branes are topological anti-branes, colors for the holonomy need to be transposed to the one for the ordinary branes. In addition, the role of the equivariant parameters should be exchanged $(q,t) \leftrightarrow (t^{-1},q^{-1})$ due to \eqref{omega} for the holonomy of the $\bar t$-branes. Thus, the partition function for the $\bar t$-branes takes the form
\begin{equation}\label{def-t}
Z_{\defo,\SU(N)}^{\bar t}=\sum_\lambda  \overline \rCS_{\SU(N),\lambda}(T_{m,n}; q,t) \;  P_{\lambda^T}(-x;t,q) ~,
\end{equation}
where we use the property $P_{\lambda^T}(-x;t^{-1},q^{-1})= P_{\lambda^T}(-x;t,q)$ of the Macdonald functions presented in \eqref{mac-inverse}.

\section{Resolved conifold side}\label{sec:resolved}

In the seminal papers  \cite{Gopakumar:1998jq,Gopakumar:1998ii}, Gopakumar and Vafa  (GV) proposed that an effective action of Type IIA string theory compactified on a Calabi-Yau manifold can be determined by considering effects of BPS particles arising M2-branes in the anti-selfdual graviphoton background. Moreover, the form of an effective action has been explicitly evaluated by applying Schwinger computations. Subsequently, Labastida, Mari\~no, Ooguri and Vafa (LMOV)  \cite{Ooguri:1999bv,Labastida:2000zp,Labastida:2000yw,Labastida:2001ts} have carried out similar analyses in the presence of D4-branes, which can be regarded as an open-string analogue of  the GV formula. At the unrefined level, thorough analysis and elaborate explanation for these formulas have been presented in \cite{marino2005chern,Dedushenko:2014nya}.  In this section, we will find the explicit form of an effective action of Type IIA string theory with D4-branes in the  \emph{refined} case.

\subsection{Without M5-branes}\label{sec:closed}

To this end, let us first review an effective action of Type IIA string theory compactified on a Calabi-Yau manifold $X$ without D4-branes. In this subsection, we assume that a Calabi-Yau three-fold $X$ is general and we do not necessarily restrict ourselves to the resolved conifold. The effective action (free energy) of Type IIA string theory has F-terms that admit genus expansion
$$
F^{\textrm{cl}}=\log Z^{\textrm{cl}}_{{\textrm{res}}}=-i\sum_{g\ge0}\int_{\bR^4} \dd^4x\, \dd^4\theta\; \cF_g(\cX_\Lambda) (\cW_{AB}\cW^{AB})^g~,
$$
where $\cF_g$ are holomorphic functions of chiral superfields  $\cX_\Lambda$ ($\Lambda=0,\cdots, b_2(X))$ associated to vector multiplets and $\cW_{AB}$ is a chiral superfield whose bottom component is anti-selfdual graviphoton field strength
$$
\cW_{AB}=\frac12 T_{AB} -R_{ABCD} \theta \sigma^{CD} \theta+ \cdots~.
$$
Since the graviphoton field strength takes the value $T=g_{\textrm{st}}(\dd x_1\wedge \dd x_2 -\dd x_3\wedge \dd x_4)$, the genus $g$ contribution is proportional to $g_{\textrm{st}}^{2g-2}$.

Gopakumar and Vafa proposed that, instead of evaluating each genus amplitude $\cF_g$ in Type IIA string theory, the whole effective action $F^{\textrm{cl}}$ can be obtained by summing up contributions of BPS states in  M-theory. They arise from M2-branes  wrapped on a holomorphic curve $\Sigma$ in the Calabi-Yau manifold $X$. This calculation can be considered as a supersymmetric version of Schwinger's computation of a one-loop effective action due to a charged particle in a constant magnetic field.

To see the spin content of BPS states, let us recall  the five-dimensional $\mathcal{N}=1$ supersymmetry algebra consisting of eight supercharges with $\Sp(1)_{R}=\SU(2)_{R}$ $\cR$-symmetry.  Since the graviphoton field breaks the five-dimensional Lorentz group $\SO(1,4)$, it is convenient to rewrite the algebra in terms of four-dimensional notation. Then, the supercharges can be organized as $Q^{I}_{\alpha}, Q^{I}_{\dot{\alpha}}$
where $I=1,2$ are the $\SU(2)_{R}$ indices, and $\alpha$ and $\dot{\alpha}$ are negative and positive chirality of the rotational group $\SO(4)=\SU(2)_{\ell}\times \SU(2)_{r}$.  With this notation, the supersymmetry algebra is given by
\begin{align}\nonumber
\{Q_{\a}^I,Q_{\b}^J\}&=\varepsilon_{\a\b}\varepsilon^{IJ}( H+\zeta)~, \cr
 \{Q_{\dot \alpha}^I, Q_{\dot \beta}^J\}&=\varepsilon_{\dot \alpha\dot \beta}\varepsilon^{IJ}(H-\zeta)~, \cr
 \{Q_{\a}^I,Q_{\dot \beta}^J\}&= -i\Gamma^\mu_{\alpha\dot \beta}\varepsilon^{IJ}P_\mu~.
 \end{align}
where $\zeta$ is the \emph{real} five-dimensional central charge.  Thus, short left-handed BPS multiplets take the form,
\begin{equation}\label{left-BPS}
 \Big((0,0;\tfrac{1}{2})\oplus (\tfrac{1}{2},0;0)\Big) \otimes(J_\ell,J_{r};S_{R})~,
\end{equation}
which represent BPS particles of mass $m = \zeta$ at rest. The first spin content is indeed the half-hypermultiplet representation and the second is  an arbitrary finite dimensional representation of $\SU(2)_\ell\times \SU(2)_r\times \SU(2)_R$. It is easy to see that the unrefined index
$$
\mathrm{Tr}(-1)^{2(J_\ell+J_{r})}q^{2J_\ell}e^{-\beta H}
$$
receives contributions only from the short left-handed BPS multiplets. Hence, a one-loop calculation involving small fluctuations around a BPS particle trajectory takes the form
\begin{align}
F^{\textrm{cl}}&=-\int_0^{\infty} \frac{ds}{s}\frac{\textrm{Tr}_\mathcal{BPS}(-1)^{2(J_\ell+J_{r})}q^{2sJ_\ell}
e^{-s m} }{(q^{\frac{s}{2}}-q^{-\frac{s}{2}})^2}\ ,\cr
&=-\sum_{d>0}\sum_{\beta \in H_2(X,\bZ)}\frac1d\frac{\textrm{Tr}_\mathcal{\cH(\beta)}(-1)^{2(J_\ell+J_{r})}q^{2dJ_\ell}}{(q^{\frac{d}{2}}-q^{-\frac{d}{2}})^2}e^{-d\beta \cdot \tau}~,
\label{Gopakumar-Vafa}
\end{align}
where we identify the parameters by $q=e^{ig_{\textrm st}}$ and $m$ is the central charge of BPS particles. Let us closely look at the meaning of the GV formula \eqref{Gopakumar-Vafa}.
\begin{itemize}
\item The denominator is obtained by Schwinger computation for the one-loop determinant of BPS particle with the half-hypermultiplet representation in the anti-selfdual graviphoton background of the form
\be \label{graviphoton}
T^-=\frac12\left(\begin{array}{cccc}
0 &g_{\textrm{st}}&&\\
-g_{\textrm{st}}&0&&\\
&&0&-g_{\textrm{st}}\\
&&g_{\textrm{st}}&0
\end{array}\right)~.
\ee
The two-dimensional contribution (the upper block) can be evaluated by summing up all the Landau levels $(\frac{1}{2}+n)g_{\textrm{st}}$ for $n\in\bZ_{\ge0}$:
$$
\sum_{n\ge0} e^{-i g_{\textrm{st}}(1+2n)/2}=\frac{1}{q^{\frac12}-q^{-\frac{1}{2}}}~.
$$
Including an identical factor for the lower block, the Schwinger calculation provides the denominator.
\item The unrefined index in the numerator receives contributions from other massive BPS states \eqref{left-BPS}, which can be understood as fermion zero modes on M2-branes. The detail analysis for the BPS spectra will follow below.
\item The central charge of a BPS state is given by the area of a holomorphic curve $\Sigma$ the M2-brane wraps and the Kaluza-Klein momentum. If we take a basis $C_I$ ($I=1,\cdots,b_2(X)$) of $H_2(X,\bZ)$, then the homology class of the curve is expressed by $[\Sigma]=\sum_I\beta_IC_I$ with $\beta_I\in \bZ$.
Denoting the complexified Kahler parameter of the 2-cycle $C_I$ by $\tau^I$, the area of the M2-brane is equal to $\tau \cdot \beta=\sum_I \tau^I \beta_I$. Then, the central charge of the BPS state is given by $m = \tau \cdot \beta + 2\pi i n $ where  $n$ is the Kaluza-Klein momentum of the M2-brane along the M-theory circle.
\item From the first to the second line, we perform a Poisson resummation $\sum_{n\in \bZ}e^{-2\pi i n s}=\sum_{d\in\bZ}\delta(s-d)$, which re-expresses the sum over the Kaluza-Klein momenta as a sum over winding numbers.
\end{itemize}

In refined topological string theory, the graviphoton field is no longer anti-selfdual, and it rather takes the form
\be\label{graviphoton-refine}
T=\e_1 dx_1\wedge dx_2-\e_2 dx_3\wedge dx_4~,
\ee
which introduces the $\Omega$-background \eqref{omega}. Furthermore,
by using the $\SU(2)_R$ $\cR$-symmetry, we define a refined index  \cite{Nekrasov:2002qd} as
$$
\mathrm{Tr}(-1)^{2(J_\ell+J_{r})}q_{\ell}^{2J_\ell}q_{r}^{2(J_{r}-S_{R})}e^{-\beta H}~.
$$
Then, one can confirm that only left-handed multiplets \eqref{left-BPS} again contribute to the index while the long multiplets and the right-handed multiplets do not. To write the refined index in terms of the equivariant parameters $q=e^{i\e_1}$ and $t=e^{i\e_2}$, we introduce $J_1=J_{\ell}+J_{r}$ and $J_{2}= J_{r}-J_{\ell}$ so that it takes the form
$$
\mathrm{Tr} (-1)^{F}q^{J_{1}-S_{R}}t^{S_{R}-J_{2}}e^{-\beta H}~,
$$
where we define $q_\ell=(qt)^{1/2}$ and $q_r=(q/t)^{1/2}$.
Then, the free energy at the refined level can be written as
\begin{align}
F_{\textrm{ref}}^{\textrm{cl}}&=\int_0^{\infty} \frac{ds}{s}\frac{\textrm{Tr}_\mathcal{BPS}(-1)^{2(J_\ell+J_{r})}q^{s(J_{1}-S_{R})}t^{s(S_{R}-J_{2})}
e^{-s m} }{(q^{\frac{s}{2}}-q^{-\frac{s}{2}})(t^{-\frac{s}{2}}-t^{\frac{s}{2}})}\ ,\cr
&=\sum_{d>0}\sum_{\beta \in H_2(X,\bZ)}\frac1d\frac{\textrm{Tr}_\mathcal{\cH(\beta)}(-1)^{2(J_\ell+J_{r})}q^{d(J_{1}-S_{R})}t^{d(S_{R}-J_{2})} }{(q^{\frac{d}{2}}-q^{-\frac{d}{2}})(t^{-\frac{d}{2}}-t^{\frac{d}{2}})}e^{-d\beta \cdot \tau}~.
\nonumber
\end{align}
Since we now have different equivariant parameters the upper and lower block in \eqref{graviphoton}, the denominator is resolved as $(q^{\frac{1}{2}}-q^{-\frac{1}{2}})^2 \to (q^{\frac{1}{2}}-q^{-\frac{1}{2}})(t^{\frac{1}{2}}-t^{-\frac{1}{2}})$.

Now let us study other massive BPS states that contribute to the refined index. A propagating M2-brane wrapped on a holomorphic curve $\Sigma_g\subset X$ of genus $g$ generates a 5d particle that preserves a half of superysymmetry \cite{Becker:1995kb}. For the low energy description, we need to take into account fermion zero-modes on $\Sigma_g$ where half of them transform under $\SU(2)_\ell \times\SU(2)_r$ as $(\frac12,0)$ and the other half transform as $(0,\frac12)$. The $(\frac12,0)$  fermionic zero modes can be interpreted as differential forms on $\Sigma_g$ and therefore there are zero-modes consisting of $2g$ copies of the $(\frac12,0)$ representation on the curve $\Sigma_g$ of genus $g$.
In other words, the $(\frac12,0)$ zero modes on $\Sigma_g$ can be interpreted as the cohomology of the Jacobian of $\Sigma_g$ where  the $\SU(2)_\ell$ action can be understood as the natural Lefschetz $\SU(2)$ action on the cohomology $\cH_g:=H^*(\textrm{Jac}(\Sigma_g))$ of the Jacobian. Quantization of this system  gives $g$ copies of the spin content $(0,0;\tfrac{1}{2})\oplus (\tfrac{1}{2},0;0)$ under $\SU(2)_\ell\times\SU(2)_r\times\SU(2)_R$ so that the contribution to the index is
$
(q^{\frac{1}{2}}-q^{-\frac{1}{2}})^g(t^{-\frac{1}{2}}-t^{\frac{1}{2}})^g~.
$

On the other hand, the $(0,\frac12)$ fermion zero modes are related by supersymmetry to infinitesimal deformations of $\Sigma_g$ as a holomorphic curve in $X$.   Let us denote a moduli space $\cM_{g,\beta}$ that parametrizes the holomorphic deformations of $\Sigma_g$ under the homology class $\beta\in H_2(X,\bZ)$ inside the Calabi-Yau three-fold $X$. Then, if $\cM_{g,\beta}$ and $\Sigma_g$ are both smooth, the total moduli space $\widehat \cM_{\textrm{cl}}$ for the BPS states in this configuration is
\be\label{fibration-closed}\begin{tikzcd}
\textrm{Jac}(\Sigma_g) \arrow[r]
& \widehat \cM_{\textrm{cl}} \arrow[d, "\pi" ] \\
& \cM_{g,\beta}
\end{tikzcd}\qquad.
\ee
Since the zero-modes of $(0,\frac12)$ fermions are differential forms on $\cM_{g,\beta}$, the space of total BPS states in this situation is the de Rham cohomology $H^*(\cM_{g,\beta};\cH_g)$ of $\cM_{g,\beta}$ with values in $\cH_g$ where the $\SU(2)_r$ action is the natural Lefschetz $\SU(2)$ action on $H^*(\cM_{g,\beta})$.
However, the assumption that $\cM_{g,\beta}$ and $\Sigma_g$ are always smooth is almost never satisfied and there are usually singular fibers in \eqref{fibration-closed}, which makes it difficult to give a rigorous mathematical definition of GV invariants. (However, important progress has been made recently in mathematics \cite{pandharipande2010stable,Maulik:2016rip}.)

In physics, one can formally count the number of BPS states. For this purpose, we decompose the zero-modes of $(0,\frac12)$ fermions into the spectrum $H^*(\cM_{g,\beta})\cong\bigoplus A_{g,\beta,J_r,S_R}$ with respect to $J_r$ and $S_R$ spins so that the total
BPS spectrum takes the form $$\cH(\beta)\cong \bigoplus_{g,J_r,S_R} \cH_g \otimes A_{g,\beta,J_r,S_R}~. $$ As a result, denoting the number of states with fixed charges by  $$\wh N_{g,\beta,J_r,S_R}:=\dim A_{g,\beta,J_r,S_R}~,$$ the refined free energy takes the form
$$
F_{\textrm{ref}}^{\textrm{cl}}=\sum_{d>0}\sum_{\textrm{charges}}\frac1d (-1)^{2J_r}\wh N_{g,\beta,J_r,S_R} (q^{\frac{d}{2}}-q^{-\frac{d}{2}})^{g-1}(t^{-\frac{d}{2}}-t^{\frac{d}{2}})^{g-1}e^{-d\beta\cdot \tau} \sum_{j_r=-J_r}^{J_r} \sum_{s_R=-S_R}^{S_R} \left(\frac{q}{t} \right)^{d(j_r-s_R)}~,
$$
where charges are summed over $g\ge0$, $\beta\in H_2(X,\bZ)$, and $J_r,S_R\in \frac12 \bZ_{\ge0}$. In particular, the integral numbers for unrefined BPS states
$$
n_{g,\beta}=\sum_{J_r,S_R\in \frac12 \bZ_{\ge0}}(-1)^{2J_r}(2J_r+1)(2S_R+1)\wh N_{g,\beta,J_r,S_R}
$$
are called GV invariants, and $n_{g,\beta}=(-1)^{\dim_\bC \cM_{g,\beta}}\chi(\cM_{g,\beta})$ if $\cM_{g,\beta}$ is smooth.
We refer the reader to \cite[references therein]{Aganagic:2012hs,Choi:2012jz,Chuang:2013wpa,Nekrasov:2014nea,Gu:2017ccq} for recent developments on refined closed BPS invariants.

\subsection{With M5-branes}\label{sec:open}

Now let us include M5'-branes on $S^1\times \bR^2\times \ccL\subset S^1\times \bR^4\times X$ like \eqref{resolved} where $\ccL$ is a special Lagrangian submanifold of a Calabi-Yau three-fold $X$. In general, a half of supersymmetry is preserved even if we include a number of M5'-branes unless their supports $\ccL_i\subset X$ are special Lagrangian \cite{Becker:1995kb}. By reducing on the M-theory circle, we have Type IIA string theory with D4-branes. Then, its low energy effective action has terms that are supported on the world-volume $\bR^2\subset\bR^4$ of the D4-branes so that it takes form
\begin{equation}\label{free-open}
F^{\textrm{op}}=\sum_{g,h\ge0}\int_{\bR^4} \dd^4x\, \dd^4\theta\; \delta(x^2) \delta(\theta^2) \; \cF_{g,h}(\cX_\Lambda;\cV_\sigma)(\cW^2)^g\cW_\parallel^{h-1}~,
\end{equation}
where  $\cV_\sigma$ ($\sigma=1,\dots,
 b_1(\ccL)$) are chiral superfields associated to the moduli of $\ccL$.
 Here $\cW_\parallel$ is the ``parallel'' component of the graviphoton superfield $\cW_{ A B}$. More precisely, for the anti-selfdual graviphoton background \eqref{graviphoton}, $T^-_{12}=T_\parallel/2$ if the D4-branes are located on the  $D_q$ plane and $T^-_{34}=T_\parallel/2$ if they are put on the $D_{\bar t}$ plane where $T_\parallel$ is  the bottom component of the superfield $\cW_\parallel$. At the unrefined level, each integral in \eqref{free-open} is proportional to $g_s^{2g+h-2}=g_s^{-\chi}$, which is natural from the viewpoint of string perturbation theory since the Euler characteristic of a Riemann surface of genus $g$ with $h$ holes is equal to $\chi=2-2g-h$.

Even in the presence of M5'-branes, one can apply the same idea that $F^{\textrm{op}}$ can be determined by analyzing contributions of BPS states in M-theory \cite{Ooguri:1999bv,Labastida:2000zp,Labastida:2000yw,Labastida:2001ts}.  A relevant BPS state arises from an M2-brane wrapped on $\Sigma\subset X$ and generally attached to the M5'-branes on $\ccL$ so that they propagate only along the M5'-branes.   From a low energy point of view, these states propagate along $S^1 \times\bR^2\subset S^1 \times\bR^4$.

To evaluate contributions to the Type IIA effective action (free energy) supported on $\bR^2$, let us investigate quantum numbers of BPS states in three-dimensional $\mathcal{N}=2$ supersymmetric theory. The three-dimensional $\mathcal{N}=2$ supersymmetry algebra is given by
\begin{align}\nonumber
&\{Q_{\alpha},\overline{Q}_{\beta}\}  =  -i\sigma^{\mu}_{\alpha\beta}P_{\mu} + i\epsilon_{\alpha\beta}\zeta~, \cr
&\{Q_{\alpha},Q_{\beta}\}  =  0 =\{\overline{Q}_{\alpha}, \overline{Q}_{\beta}\} ~.
\end{align}
For a theory on the $q$-branes, the supercharges $Q$ are complex spinors in the spin-$\frac12$ representation of the rotation group $\SO(3)\cong\SU(2)_{1}$, which is the diagonal subgroup of $\SU(2)_\ell\times \SU(2)_r$. In addition,  a $\U(1)_2$ $\cR$-symmetry, which is actually the $\U(1)$ subgroup of  the anti-diagonal subgroup $\SU(2)_2=\{(x,x^{-1})\in \SU(2)_\ell\times\SU(2)_r\}$ in five-dimension, rotates $Q_{\alpha}$ and $\overline{Q}_{\alpha}$. (See Table \ref{tab:supercharge}.) For a theory on the $\bar t$-branes, the roles of $\SU(2)_1$ and $\SU(2)_2$ are exchanged so that the rotational group is identified with $\SU(2)_2$ and the $\cR$-symmetry is $\U(1)_1\subset \SU(2)_1$.

The three-dimensional unrefined index defined by
\begin{equation}\label{open-unrefine}
\textrm{Tr}(-1)^{F}q^{J_1-J_2}e^{-\beta H}~
\end{equation}
counts states annihilated by the supercharges  $Q_{+}$ and $\overline  Q_{-}$.  Then, only the left short multiplets, which represent BPS particles of mass $M = \zeta$ at rest,
\be\label{spin-open}
\left((0;-\tfrac12)\oplus(-\tfrac12;0)\right) \otimes (J_1;J_2)~,
\ee
contribute to the index. The proposal of LMOV is that the free energy \eqref{free-open} in the anti-selfdual graviphoton background \eqref{graviphoton} takes the form
\begin{align}\nonumber
F^{\textrm{op}}&=\pm\int_0^{\infty} \frac{ds}{s}\frac{\textrm{Tr}_\mathcal{BPS}(-1)^{F}q^{s(J_{1}-J_{2})}
e^{-s m} }{(q^{\frac{s}{2}}-q^{-\frac{s}{2}})}\ ,\cr
&=\pm\sum_{d>0}\sum_{\beta \in H_2(X,\bZ)}\sum_{\vec k}\frac1d\frac{\textrm{Tr}_{\cH(\beta,\vec{k})}(-1)^{F}q^{d(J_{1}-J_{2})}}{(q^{\frac{d}{2}}-q^{-\frac{d}{2}})}e^{-d\beta \cdot \tau}p_{\vec k}(x)~.
\end{align}
\begin{itemize}
\item Since BPS particles propagate only along the M5'-branes, the Schwinger computation is here performed only on $\bR^2\subset \bR^4$ spanned by the M5'-branes so that the denominator originates from either the upper ($q$-brane) or the lower block (${\bar t}$-brane) of \eqref{graviphoton}, depending on the M5'-brane configurations.
\item As in the closed case \S \ref{sec:closed}, the unrefined index in the numerator receives contributions from other massive BPS states \eqref{spin-open}, which can be understood as fermion zero modes on M2-branes attached to the M5'-branes. The detail analysis for the BPS spectra will be given below.
\item The central charge of a BPS state is expressed by the area of the M2-brane as well as the Kaluza-Klein modes. The area of a holomorphic curve $\Sigma\subset X$ whose boundary is on $\ccL$ is determined by its relative homology class in $H_2(X,\ccL;\bZ)$. A Lagrangian subvariety $\ccL$ in \eqref{resolved} for a knot is topologically homeomorphic to $S^1\times \bR^2$ \cite{harvey1982calibrated,Aganagic:2000gs}, which simplifies the relative homology as $$H_2(X,\ccL;\bZ)\cong H_2(X;\bZ)\oplus H_1(\ccL;\bZ)~.$$
Hence, the homology class $[\Sigma]\in H_2(X,\ccL;\bZ)$ is expressed by $\beta=(\beta_1,\cdots,\beta_{b_2(X)}) \in H_2(X;\bZ)$ as in the closed string as well as the winding numbers $w=(w_1,\cdots,w_h) \in (H_1(\ccL;\bZ))^h\cong\bZ^h$ of boundary components $\partial \Sigma\cong (S^1)^h$. Since $\Sigma$ is oriented, one can assume that $w_i$ are all non-negative integers.  To express the contribution from the boundary components, we define a vector $\vec k$ as follows: the $i$-th entry of $\vec k$ is the number of $w_i$'s that take the value $i$.  Then, when $M$ M5'-branes wrap on $S^1\times\bR^2\times \ccL$, we can write
$$
e^{-m} = e^{-\tau \cdot \beta - 2\pi i n } ~p_{\vec k}(x)~,
$$
where the fugacities $x$ parametrize the Cartan subgroup of the $\U(M)$-valued moduli of $\ccL$ and $p_{\vec k}(x)$ is defined in Appendix \S\ref{sec:app-a}. Let us note that it is easy to transform from the winding basis $p_{\vec k}(x)$ to the representation basis $s_\mu(x)$ for the moduli of $\ccL$ by using \eqref{Schur-Newton}.

\end{itemize}

\begin{wraptable}{L}{0.4\textwidth}
\centering
\begin{tabular}[c]{| c | | c | c | c |}
\hline
& $2J_1$ & $2J_2$ & $2S_R$ \\
\hline
$Q_{+}$ & $+1$ & $+1$ & $+1$ \\
\hline
$Q_{-}$ & $-1$ & $+1$ & $+1$ \\
\hline
$\overline Q_{+}$ & $+1$ & $-1$ & $-1$ \\
\hline
$\overline  Q_{-}$ & $-1$ & $-1$ & $-1$ \\
\hline
\end{tabular}
\caption{Charges for 3d $\mathcal{N}=2$ supersymmetry on the $q$-brane.}\label{tab:supercharge}
\end{wraptable}
Now, let us consider the case where $X$ is the resolved conifold and $\ccL$ is the configuration $\ccL_{T_{m,n}}$ for a torus knot as in \eqref{resolved}.
As we have discussed in \S\ref{sec:deformed}, this configuration preserves the extra $\U(1)_R$ global symmetry, which can be actually interpreted as the $\U(1)$ subgroup of the $\SU(2)_R$ $\cR$-symmetry in five dimension with eight supercharges.  In this case, the three-dimensional $\cN=2$ supercharges have the quantum numbers under this symmetry shown in Table \ref{tab:supercharge}. Since $J_1-S_R$ and $S_R-{J}_2$ commute with the supercharges $Q_{+}$ and $\overline  Q_{-}$, one can refine the index by
$$
{\rm Tr}\, (-1)^{F}\, q^{J_1-S_R}\, t^{S_R-{J}_2}e^{-\beta H}~.
$$

Thus, one can turn on the $\Omega$-background \eqref{graviphoton-refine} in the presence of M5'-branes supported on $S^1\times \bR^2\times \ccL_{T_{m,n}}$ by using the refined index. First, we shall consider the $q$-brane setting in \eqref{resolved} at the refined level. In the refined graviphoton background \eqref{graviphoton-refine}, the Schwinger computation provides the form of the free energy
$$
F_{\textrm{ref}}^{q}=\sum_{d>0}\sum_{\beta \in H_2(X,\bZ)}\sum_{\mu}\frac1d\frac{\textrm{Tr}_\mathcal{\cH(\beta,\mu)}(-1)^{F}q^{d(J_{1}-S_{R})}t^{d(S_{R}-J_{2})} }{q^{\frac{d}{2}}-q^{-\frac{d}{2}}}e^{-d\beta \cdot \tau}s_\mu(x^d)~.
$$

The BPS states that contribute to the refined index are fermion zero modes on an M2-brane wrapped on a holomorphic curve $\Sigma_{g,h}\subset X$ whose boundary is on $\ccL$. Since the presence of the M5'-branes breaks $\SU(2)_\ell\times\SU(2)_r\times \SU(2)_R$ to  $\SU(2)_1\times\U(1)_2\times \U(1)_R$, it is not so straightforward to study quantum numbers of BPS states as in \S \ref{sec:closed}. However, as before, fermion zero modes on an M2-brane can be associated to cohomology groups of the  moduli space
\be\nonumber\begin{tikzcd}
\textrm{Jac}(\Sigma_{g,h}) \arrow[r]
& \widehat \cM_{\textrm{op}} \arrow[d, "\pi" ] \\
& \cM_{g,h,\beta}
\end{tikzcd}\qquad,
\ee
where  the moduli space $\cM_{g,h,\beta}$ parametrizes deformations of $\Sigma_{g,h}\subset X$ that preserve a half of supersymmetry. More precisely, as emphasized in \cite{Labastida:2000yw}, the fermion zero modes are the cohomology groups $H^*(\widehat \cM_{\textrm{op}})\cong   H^*(\textrm{Jac}(\Sigma_{g,h}))
\otimes H^*({\cal M}_{g,h,\beta})$ of the moduli space \emph{mod out by the action (Sprecht module) of the permutation group} $\frakS_h$ where the group $\frakS_h$ exchanges $h$  distinguished holes of $\Sigma_{g,h}$. The Jacobian $\textrm{Jac}(\Sigma_{g,h})$ of a curve of genus $g$ with $h$ holes is topologically $ (T^{2})^g\times (S^1)^{h-1}$ where the permutation group $\frakS_h$ does not act on the cohomology $H^*((T^{2})^g)$ of the Jacobian of the ``bulk'' Riemann surface. Therefore, the contribution to the refined index from $H^*((T^{2})^g)$ is $
(q^{\frac{1}{2}}-q^{-\frac{1}{2}})^g(t^{-\frac{1}{2}}-t^{\frac{1}{2}})^g$ as in \S\ref{sec:closed}.

 The projection of $H^*((S^1)^{h-1}) \otimes H^*({\cal M}_{g,h,\beta})$ onto the invariant subspace can
be done by using the Schur functor ${\bf S}_{\mu}$. More explicitly, the invariant subspace in BPS states $\cH(\beta,\mu)$ with charge $\beta\in H_2(X,\bZ)$ and $\mu$ (a representation of $\U(M)$) can be written as
\begin{multline}\label{knot-BPS}
{\rm Inv}\Bigl( H^*((S^1)^{h-1})
\otimes H^*({\cal M}_{g,h,\beta})\Bigr)_{\cH(\beta,\mu)}=\\
\bigoplus_{\sigma,\rho} C_{\mu\sigma\rho}
{\bf S}_{\sigma}(H^*((S^1)^{h-1}))
\otimes {\bf S}_{\rho}(H^*({\cal M}_{g,h,\beta}))~,
\end{multline}
where the Clebsch-Gordon coefficients
$C_{\mu\sigma\rho}$ of the permutation group $\frakS_{h}$ are
\begin{align}\label{CG}
C_{\mu\sigma\rho} =\sum_{\vec k} {|C(\vec k)| \over k !}
\chi_\mu (C(\vec k))  \chi_{\sigma} (C(\vec k)) \chi_{\rho} (C(\vec k)) ~.
\end{align}
Indeed, they are symmetric under the permutations of $({\mu,\sigma,\rho})$.

The cohomology of the Jacobian of $\Sigma_{g,h}$ can be interpreted as differential forms on $\Sigma_{g,h}$. In particular, the boundary part $H^*((S^1)^{h-1})$ is spanned by one-forms $d\theta_i$, ($i=1, \cdots, h$), which are
Poincar\'e dual to the holes in the curve $\Sigma_{g,h}$, and they are subject to the linear constraint $\sum_i d\theta_i=0$. Moreover, one can consider the differential form $d\theta_i$ as the fermion zero modes $\psi_i$ on a rigid curve with charges
\be\label{fermion-cherge}
(J_1;J_2,S_R)=(\tfrac12;-\tfrac12,\tfrac12)~.
\ee
As explained in \cite{Labastida:2000yw,marino2005chern}, the BPS spectra ${\bf S}_{\sigma}(H^*((S^1)^{h-1}))$ can be obtained by acting the fermion zero modes $\psi_i$ on the vacuum $|0\rangle$. The defining representation $V$ of $\frakS_h$ can be constructed by acting one fermion $\psi_i$ on the vacuum $|0\rangle$, and its dimension is $h-1$ due to $\sum_i\psi_i=0$. The rest of the spectra are generated by taking the wedge products $\wedge^d V$ of the defining representation. Assigning the Young diagram $\yng(9)$ with $h$ boxes of one row to the trivial representation $|0\rangle$, the irreducible representations $\wedge^d V$ of $\frakS_h$ are called \emph{hook representations} since their Young tableau are of the form with ($h-d$)-boxes in the first row
$$
\yng(6,1,1,1,1)~.
$$
Assuming that the vacuum $|0\rangle$ is neutral, the state $\wedge^d V$ has charges
$$
(J_1;J_2,S_R)=(\tfrac{d}{2};-\tfrac{d}{2},\tfrac{d}{2})~,
$$
because it is essentially obtained by acting $d$-wedge products of the fermions charged by \eqref{fermion-cherge}. Hence, the contribution from the state $\wedge^d V$ to the refined index is $(-t)^d$. As a result, the refined index only over the BPS states ${\bf S}_{\sigma}(H^*((S^1)^{h-1}))$
$$
B_\sigma:={\rm Tr}_{{\bf S}_{\sigma}(H^*((S^1)^{h-1}))}\, (-1)^{F}\, q^{J_1-S_R}\, t^{S_R-{J}_2}~
$$
is summarized as
\begin{equation}
B_\sigma(t)= \left\{
\begin{array}{ll}
(-t)^d t^{-\frac{|\sigma|-1}{2}} &\qquad \sigma: \mathrm{hook\ rep\ for} \wedge^d V \\
0 &\qquad\sigma: \mathrm{otherwise}
\end{array}
\right.\nonumber~~.
\end{equation}
In fact, we normalize $B_\sigma$ by $ t^{-\frac{|\sigma|-1}{2}}$ so that they satisfy
\be\label{B-mirror}
B_\sigma(t^{-1}) = (-1)^{|\sigma|-1}B_{\sigma^T} (t)~.
\ee

As it can be easily seen, the BPS states $\cH_{\sigma,g}\cong H^*((T^{2})^g)\otimes {\bf S}_{\sigma}(H^*((S^1)^{h-1}))$ obey $J_1+J_2=0$, corresponding to the $(\tfrac12,0)$ fermions under $\SU(2)_\ell\times\SU(2)_r$ in five dimension. On the other hand, the BPS states ${\bf S}_{\rho}(H^*({\cal M}_{g,h,\beta}))$ satisfy $J_1-J_2=0$, analogous to the $(0,\tfrac12)$ fermions in five dimension. Although they can contribute to the unrefined index \eqref{open-unrefine} only by signs, the refined index receives non-trivial contributions. Defining $J_r:=\frac12(J_1+J_2)$, one can decompose ${\bf S}_{\rho}(H^*({\cal M}_{g,h,\beta}))$ into the spectrum $\bigoplus A_{\rho,g,\beta,J_r,S_R}$ with respect to $J_r$ and $S_R$ charges so that the total BPS states are
$$
\cH(\beta,\mu)\cong \bigoplus_{\sigma,\rho,g,J_r,S_R} C_{\mu\sigma\rho}\;\cH_{\sigma,g}\otimes A_{\rho,g,\beta,J_r,S_R}~.
$$
 As a result, writing the number of BPS states with fixed charges by
 \be\label{hat-N}
 \wh N_{\rho,g,\beta,J_r,S_R}:=\dim A_{\rho,g,\beta,J_r,S_R}~,
 \ee
 the refined free energy takes the form
\begin{align}\label{free-q}
F_{\textrm{ref}}^q&=\sum_{d>0}\sum_{\mu}\frac1d \frac{f_\mu^q(a^d,q^d,t^d)}{q^{\frac{d}{2}}-q^{-\frac{d}{2}}}s_\mu(x^d)~,\\
f_\mu^q(a,q,t)&=\sum_{\textrm{charges}} (-1)^{2J_r} C_{\mu\sigma\rho}B_{\sigma}(t)  \wh N_{\rho, g,\beta,J_r,S_R}(q^{\frac{1}{2}}-q^{-\frac{1}{2}})^{g}(t^{-\frac{1}{2}}-t^{\frac{1}{2}})^{g}  \left(\frac{q}{t} \right)^{J_r-S_R-\frac\beta2}a^{\beta}~,\nonumber
\end{align}
where charges are summed over $g\ge0$, $\beta\in H_2(X,\bZ)$, $J_r,S_R\in \frac12 \bZ$, and all representations $\sigma, \rho$ of $\U(M)$. Note that, to see relation to refined Chern-Simons invariants in the next section, here we define the parameter by
$$
a:=e^{-\tau}\sqrt{\frac{q}{t}}~.
$$

In fact, the integral numbers for unrefined BPS states
\be\label{LMOV-inv}
\wh  N_{\mu,g,\beta}=\sum_{J_r,S_R\in \frac12 \bZ}(-1)^{2J_r}\wh  N_{\mu,g,\beta,J_r,S_R} ~
\ee
are called LMOV invariants. Moreover,  Mari\~no and Vafa proposed the multi-covering formula that relates the LMOV invariants ${\widehat N}_{\rho,g,\beta}$ to open Gromov-Witten invariants in the presence of D4-branes supported on $\bR^2\times \ccL$ \cite{Marino:2001re}. In the case of the framed unknot, the multi-covering formula has been proven based on localization method and combinatorics \cite{Liu:2003jk}.

Next let us consider the $\bar t$-brane setting in \eqref{resolved}. The form of the free energy is
$$
F_{\textrm{ref}}^{\bar t}=\sum_{d>0}\sum_{\beta \in H_2(X,\bZ)}\sum_{\mu}\frac1d\frac{\textrm{Tr}_\mathcal{\cH(\beta,\mu)}(-1)^{F}q^{d(J_{1}-S_{R})}t^{d(S_{R}-J_{2})} }{t^{-\frac{d}{2}}-t^{\frac{d}{2}}}e^{-d\beta \cdot \tau}s_{\mu}(x^d)~,
$$
where the Schwinger computation on the $D_{\bar t}$ plane yields the denominator. Although the BPS spectra in the $\bar t $-brane setting are essentially  the same as those in the $q$-brane setting, their $J_1$ and $J_2$ charges are exchanged. Hence, the fermion zero modes $\psi_i$ for $H^*((S^1)^{h-1})$ have charges $(J_2;J_1,S_R)=(\frac12;-\frac12,\frac12)$ so that the refined index over the BPS states ${\bf S}_{\sigma}(H^*((S^1)^{h-1}))$ is given by $B_\sigma(q^{-1})$. The other part is exactly the same as the $q$-brane setting so that the free energy for the $\bar t$-branes is
\begin{align}\label{free-t}
F_{\textrm{ref}}^{\bar t}&=\sum_{d>0}\sum_{\mu}\frac1d \frac{f_\mu^{\bar t}(a^d,q^d,t^d)}{t^{-\frac{d}{2}}-t^{\frac{d}{2}}}s_{\mu}(x^d)~,\\
f_\mu^{\bar t}(a,q,t)&=\sum_{\textrm{charges}} (-1)^{2J_r} C_{\mu\sigma\rho}B_{\sigma}(q^{-1})  \wh N_{\rho, g,\beta,J_r,S_R}(q^{\frac{1}{2}}-q^{-\frac{1}{2}})^{g}(t^{-\frac{1}{2}}-t^{\frac{1}{2}})^{g}  \left(\frac{q}{t} \right)^{J_r-S_R-\frac\beta2}a^{\beta}~.\nonumber
\end{align}
It is easy to see that the free energy for the $q$-branes and that for the $\bar t$-branes are related by
\be\label{q-t}
F_{\textrm{ref}}^{\bar t}(a,q,t)=F_{\textrm{ref}}^{q}(a,t^{-1},q^{-1})~,
\ee
which can be expected from the equivariant action \eqref{omega} on $\bC^2$.

Finally, let us extract the common part of $f_\mu^{q}(a,q,t)$ and $f_\mu^{\bar t}(a,q,t)$ as
 \begin{align}\label{hat-f}
\wh f_{\rho}(a,q,t)=\sum_{\textrm{charges}}  (-1)^{2J_r}{\widehat N}_{\rho, g,\beta,J_r,S_R}  (q^{\frac12}-q^{-\frac12})^{g}(t^{-\frac12}-t^{\frac12})^{g} \left(\frac{q}{t} \right)^{J_r-S_R-\frac\beta2}a^{\beta}~~,
 \end{align}
 which is invariant under the exchange $(q,t)\leftrightarrow (t^{-1},q^{-1})$.
If we can define an invertible symmetric matrix
$$
M_{\mu\rho}(t):=
\sum_{ \sigma}
C_{\mu\sigma\rho}B_{\sigma}(t)~,
$$
then we obtain concise expressions
\begin{align}\label{f-to-hatf}
f^{q}_{\mu}(a,q,t)&= \sum_{\rho}M_{\mu\rho}(t) \wh f_{\rho}(a,q,t)~,\cr
f^{\bar t}_{\mu}(a,q,t)&=\sum_{\rho} M_{\mu\rho}(q^{-1}) \wh f_{\rho}(a,q,t)~.
 \end{align}

\section{Large $N$ duality for torus knots}\label{sec:knots}

As the number of M5-branes goes to infinity, the three-sphere $S^3$ in the deformed conifold $T^*S^3$ shrinks and the deformed conifold transforms into the resolved conifold $\cO(-1)\oplus \cO(-1)\to \bC\bP^1$.  At large $N$, the $N$ M5-branes turn into the flux supporting $\bC\bP^1$ in the resolved conifold. Thus, this duality implies that a generating function of Chern-Simons invariants of a knot at large $N$ is equal to a topological string amplitude with $M$ M5'-branes associated to the knot in the resolved conifold
$$
\lim_{\substack{N\to\infty \\ q^N=a}} Z_{\textrm{def},\U(N)}(q) = Z_{\textrm{res}}(a,q)~.
$$
The work of LMOV not only determines the form of low-energy effective actions of Type IIA string theory with D4-branes on the resolved conifold but also provides its connection to Chern-Simons invariants of a knot at large $N$. In other words, colored HOMFLY-PT polynomials are related to LMOV invariants. In this section, we shall put forth the large $N$ duality for torus knots in the refined context.

On the deformed conifold side, we have reviewed generating functions of refined Chern-Simons invariants in \S\ref{sec:deformed} . At large $N$, we substitute the stable limit  $\overline{\rCS}_{\lambda}(T_{m,n};a,q,t)$ \cite{Aganagic:2011sg,Gorsky:2013jna} for $\SU(N)$ invariants $\overline{\rCS}_{\SU(N),\lambda}(T_{m,n};q,t)$ in \eqref{def-q} and \eqref{def-t}. Then, using the forms of the refined free energy on the resolved conifold determined in \S\ref{sec:open}, the equivalences of the partition functions for both the $q$-brane and $\bar t$-brane setting can be recapitulated as
\begin{align}
\sum_{\lambda} \overline{\textrm{rCS}}_\lambda(T_{m,n};a,q,t)\; g_\lambda(q,t)P_\lambda(x;q,t) &= \exp\left( \sum_{d=1}^\infty \sum_\mu\frac{1}{d}\frac{f^{q}_\mu(T_{m,n};a^d,q^d,t^d)}{q^{\frac{d}{2}}-q^{-\frac{d}{2}}}    s_\mu(x^d)  \right)~,\label{refined-GT1}\\
\sum_{\lambda} \overline{\textrm{rCS}}_\lambda(T_{m,n};a,q,t)~   P_{\lambda^T}(-x;t,q)  &=\exp\left( \sum_{d=1}^\infty \sum_\mu\frac{1}{d}  \frac{f^{\bar t}_{\mu}(T_{m,n};a^d,q^d,t^d)}{t^{-\frac{d}{2}}-t^{\frac{d}{2}}}    s_{\mu}(x^d)  \right).\label{refined-GT2}
\end{align}
These identities determine the refined indices $f^q_\mu$, $f^{\bar t}_\mu$ and  $\wh f_\rho$ on the resolved conifold in terms of refined Chern-Simons invariants $ \overline{\textrm{rCS}}_\lambda$. Therefore, we call $f^q_\mu$, $f^{\bar t}_\mu$ and  $\wh f_\rho$ refined reformulated invariants. We shall present general formulas in Appendix \ref{sec:app-b}.

In the case of the unknot $K=\unknot$, the formulas above become the Cauchy formulas \eqref{Cauchy-mac} so that
 only two BPS numbers are non-vanishing as in the unrefined case,
\textit{i.e.} the refined reformulated invariants $f^q_{\mu}(\unknot)$, $f^{\bar t}_{\mu}(\unknot)$ colored by non-trivial representations $(\mu\neq\yng(1))$ of the unknot vanish. Geometric picture is drawn in \cite[Figure 3]{Ooguri:1999bv} where $\ccL$ is $S^1\times \bR^2$ with $S^1$ the equator of $\bC \bP^1$ in the resolved conifold and the two BPS states correspond to the M2-branes covering the upper and lower hemisphere of $\bC \bP^1$.

The refined reformulated invariants can be explicitly evaluated by using refined Chern-Simons invariants of torus knots obtained in
 \cite{Aganagic:2011sg,DuninBarkowski:2011yx,Cherednik:2011nr,Fuji:2012pm,Shakirov:2013moa}. In all the examples we have checked, the refined reformulated invariants $f^q_\mu(T_{m,n})$ and $f^{\bar t}_\mu(T_{m,n})$ obey the relation \eqref{f-to-hatf}. Moreover, after making change of variables
\begin{align}\label{cov}
a=-\mathbf{a}^2\mathbf{t}~, \quad q^{\frac12}=-\mathbf{q} \mathbf{t}~, \quad  t^{\frac12}=\mathbf{q}~,
\end{align}
the reformulated invariants $\wh f_{\rho}(T_{m,n})$ can be written in the form
 \begin{align}\label{bold-form}
\wh f_{\rho}(T_{m,n})=\sum_{g\ge 0} \sum_{\beta,F\in \bZ} \wh{\mathbf{N}}_{\rho,g,\beta,F}(T_{m,n})  (\mathbf{q}\mathbf{t}-\mathbf{q}^{-1}\mathbf{t}^{-1})^{g}(\mathbf{q}-\mathbf{q}^{-1})^{g}\mathbf{a}^{2\beta} \mathbf{t}^F~,
 \end{align}
 up on the $a$-grading shift by $\pm\frac12$.
Surprisingly, we observe that the numbers $ \wh{\mathbf{N}}_{\rho,g,\beta,F}(T_{m,n})$ are always \emph{non-negative} integers for any $\rho,g,\beta,F$. Let us emphasize that this is not obvious. Even if we assume that $\wh f_{\rho}(T_{m,n})$ takes the form \eqref{hat-f}, we have
$$
\wh{\mathbf{N}}_{\rho,g,\beta,F}(T_{m,n}) =\sum_{2(J_r-S_R)= F} (-1)^{2S_R} \wh N_{\rho, g,\beta,J_r,S_R}(T_{m,n})~,
$$
which could be negative. Since $\wh N_{\rho, g,\beta,J_r,S_R}(T_{m,n})$ are non-negative integrals by definition \eqref{hat-N}, the positivity of $\wh{\mathbf{N}}_{\rho,g,\beta,F}(T_{m,n})$ strongly suggests that the extra $\U(1)_R$ global  symmetry $S_R$ acts trivially on the BPS states ${\bf S}_{\rho}(H^*({\cal M}_{g,h,\beta}))$ so that $F$ is indeed equal to $2J_r$. The same phenomenon has been found for refinement of analytically continued WRT invariants of Lens spaces $L(p,1)$ defined by the 3d/3d correspondence \cite{Gukov:2016gkn,Gukov:2017kmk}.\footnote{S.N. would like to thank Pavel Putrov for discussion on the positivity and the action of $S_R$.}

Now let us formulate the conjecture of refined large $N$ duality for torus knots, which is the main claim of this paper.
\begin{tcolorbox}
The extra $\U(1)_R$ global symmetry $S_R$ acts trivially on the BPS states ${\bf S}_{\rho}(H^*({\cal M}_{g,h,\beta}))$ in the resolved conifold. Thus, the refined reformulated invariants $f^q_\mu(T_{m,n})$ and $f^{\bar t}_\mu(T_{m,n})$, expressed in terms of refined Chern-Simons invariants of a torus knot $T_{m,n}$ via the geometric transition \eqref{refined-GT1} and \eqref{refined-GT2} (or more explicitly  \eqref{fq-general} and \eqref{ft-general}), can be written
\begin{align}\label{f-to-hatf-2}
f^{q}_{\mu}(T_{m,n};a,q,t)&= \sum_{\rho}M_{\mu\rho}(t) \wh f_{\rho}(T_{m,n};a,q,t)~,\cr
f^{\bar t}_{\mu}(T_{m,n};a,q,t)&=\sum_{\rho} M_{\mu\rho}(q^{-1}) \wh f_{\rho}(T_{m,n};a,q,t)~,
 \end{align}
where, upon the $a$-grading shift by $\pm\frac12$,  $\wh f_{\rho}(T_{m,n})$ takes the form
  \be\label{hat-f-2}
\wh f_{\rho}(T_{m,n};a,q,t)=\sum_{\textrm{charges}}  (-1)^{2J_r}{\widehat N}_{\rho, g,\beta,J_r} (T_{m,n}) (q^{\frac12}-q^{-\frac12})^{g}(t^{-\frac12}-t^{\frac12})^{g} \left(\frac{q}{t} \right)^{J_r-\frac\beta2}a^{\beta}~~,
 \ee
 with \emph{non-negative} integers ${\widehat N}_{\rho, g,\beta,J_r} (T_{m,n})\in\bZ_{\ge0}$. Furthermore, for $\rho, g,\beta$ fixed, the $2J_r$ charges of non-zero (hence positive) integers ${\widehat N}_{\rho, g,\beta,J_r} (T_{m,n})$ are either all even or all odd so that no cancellation occurs in the unrefined limit \eqref{LMOV-inv} and therefore the LMOV invariant is
\be\label{resolution}
 \wh  N_{\rho,g,\beta}(T_{m,n})=\pm\sum_{J_r\in \frac12 \bZ}\wh  N_{\rho,g,\beta,J_r}(T_{m,n})~ .
\ee
\end{tcolorbox}

 The conjecture on the trivial action of $S_R$ implies that the numbers ${\widehat N}_{\rho, g,\beta,J_r} (T_{m,n})$ indeed yield complete information about BPS degeneracies in M-theory on the resolved conifold with the M5'-branes associated to a torus knot $T_{m,n}$. From geometric point of view, they are graded dimensions of the cohomology groups ${\bf S}_{\rho}(H^*({\cal M}_{g,h,\beta}))$ of the moduli space of M2-M5' bound states.
In Appendix \ref{sec:app-c}, we present some examples of $\wh  N_{\rho,g,\beta,J_r}(T_{m,n})$ for the trefoil $T_{2,3}$ and the $T_{2,5}$ knot. In addition, a \texttt{Mathematica} file attached to arXiv page contains more information about reformulated invariants. In all the examples, one can confirm that, for $\rho,g,\beta$ fixed, $2J_r$ charges of non-trivial ${\widehat N}_{\rho, g,\beta,J_r} (T_{m,n})$ are either all even or all odd. In addition, the property \eqref{resolution} is manifest if we compare them with tables given in \cite{Labastida:2000yw}.

It is known that refined Chern-Simons invariants $ \overline{\textrm{rCS}}_\lambda$ of a torus knot colored by non-rectangular Young diagrams generally contain both positive and negative coefficients even after the change of variables \eqref{cov}. (For instance, see \cite[\S3.4]{Cherednik:2011nr}.) However, this formulation lends itself to the natural interpretation of refined Chern-Simons invariants as a generating function of BPS states, providing non-negative integers $\wh  N_{\rho,g,\beta,J_r}(T_{m,n})$ for any color $\rho$. Thus, this can be interpreted as a \emph{positivity conjecture} of refined Chern-Simons invariants of a torus knot.

Furthermore, we notice several interesting features of refined reformulated invariants. First, instead of taking the genus expansion of M2-branes \eqref{hat-f-2}, we find that the naive change of variables \eqref{cov} for $\wh f_{\rho}(T_{m,n})$ always yields a Laurent polynomial with non-negative integral coefficients
\be\nonumber
\wh f_{\rho}(T_{m,n};a=-\mathbf{a}^2\mathbf{t},  q=\mathbf{q}^2 \mathbf{t}^2,   t=\mathbf{q}^2)=\sum_{i,j,k}\wh {\mathbf{N}}_{\mu;i,j,k}(T_{m,n})\mathbf{a}^{2i}\mathbf{q}^{2j}\mathbf{t}^k~,
\ee
 with $\wh {\mathbf{N}}_{\mu;i,j,k}\in\bZ_{\ge0}$. Hence, this evidence also indicates that there exists underlying cohomology groups of some moduli spaces for  $\wh {\mathbf{N}}_{\mu;i,j,k}$. Below some examples are given:
 \begin{align}\nonumber
& \widehat{f}_{\yng(1)}(T_{2,3})=
  \frac{\mathbf{a}^2}{\mathbf{q}^2}
  (\mathbf{a}^2 \mathbf{t}+1) (\mathbf{a}^2
   \mathbf{q}^2 \mathbf{t}^3+\mathbf{q}^4 \mathbf{t}^2+1)\cr
& \widehat{f}_{\yng(2)}(T_{2,3}) =\frac{\mathbf{a}^2}{\mathbf{q}^4} \left(\mathbf{a}^2 \mathbf{t}+1\right) \left(\mathbf{a}^2
   \mathbf{t}^3+1\right) \left(\mathbf{q}^4 \mathbf{t}^2+1\right) \left(\mathbf{a}^2
   \mathbf{t}+\mathbf{q}^2\right) \left(\mathbf{a}^2 \mathbf{q}^2
   \mathbf{t}^3+1\right)\cr
&\widehat{f}_{\yng(1,1)}(T_{2,3})\cr&= \frac{\mathbf{a}^2}{\mathbf{q}^6} \left(\mathbf{a}^2 \mathbf{t}+1\right) \left(\mathbf{a}^2
   \mathbf{t}+\mathbf{q}^2\right) \left(\mathbf{a}^2 \mathbf{q}^2 \mathbf{t}^3+1\right)
   \left(\mathbf{q}^8 \left(\mathbf{a}^2 \mathbf{t}^7+\mathbf{t}^4\right)+\mathbf{q}^4
   \mathbf{t}^2 \left(\mathbf{a}^2
   \left(\mathbf{t}^3+\mathbf{t}\right)+2\right)+\mathbf{a}^2
   \mathbf{t}^3+1\right)~.
\end{align}

Second, we also observe a positivity property when we make the same substitution \eqref{cov} for $f^{q}_{[r]}(T_{m,n})$ and $f^{\bar t}_{[r]}(T_{m,n})$ colored by symmetric representations $\lambda=[r]$:
\begin{align}
f^{q}_{[r]}(T_{m,n};-\mathbf{a}^2\mathbf{t},  \mathbf{q}^2 \mathbf{t}^2,   \mathbf{q}^2)&=\pm\mathbf{q}^{\bullet}\sum_{i,j,k} \mathbf{N}_{[r];i,j,k}^{q}(T_{m,n})\mathbf{a}^{2i}\mathbf{q}^{2j}\mathbf{t}^k~,\cr
f^{\bar t}_{[r]}(T_{m,n};-\mathbf{a}^2\mathbf{t},  \mathbf{q}^2 \mathbf{t}^2,   \mathbf{q}^2)&=\pm\mathbf{q}^{\bullet}\sum_{i,j,k}\mathbf{N}_{[r];i,j,k}^{\bar t}(T_{m,n})\mathbf{a}^{2i}\mathbf{q}^{2j}\mathbf{t}^k~,\nonumber
\end{align}
where $\mathbf{N}^{q}_{[r];i,j,k}(T_{m,n}), \mathbf{N}^{\bar t}_{[r];i,j,k}(T_{m,n})\in\bZ_{\ge0}$. For instance, we have
      \vspace{10pt}

 \noindent $\displaystyle f^q_{\yng(2)}(T_{2,3})$
     \noindent$\displaystyle =-
    \frac{\mathbf{a}^2 \mathbf{t}^2}{\mathbf{q}}$$
    (\mathbf{a}^2 \mathbf{t}+1)
   (\mathbf{q}^8 (\mathbf{a}^4 \mathbf{t}^8+\mathbf{a}^2
   \mathbf{t}^5)+\mathbf{q}^6 (\mathbf{a}^6
   \mathbf{t}^9+\mathbf{a}^4 \mathbf{t}^6+\mathbf{a}^2
   \mathbf{t}^5+\mathbf{t}^2)+\mathbf{a}^2 \mathbf{q}^4 \mathbf{t}^3
   (\mathbf{a}^2
   (\mathbf{t}^3+\mathbf{t})+2)+\mathbf{q}^2
   (\mathbf{a}^6 \mathbf{t}^5+\mathbf{a}^4 \mathbf{t}^4+\mathbf{a}^2
   \mathbf{t}+1)+\mathbf{a}^2 \mathbf{t} (\mathbf{a}^2
   \mathbf{t}+1))~,$

      \vspace{10pt}

   \noindent $\displaystyle f^{\bar t}_{\yng(2)}(T_{2,3})$
    \noindent$\displaystyle =   -
     \frac{\mathbf{a}^2}{\mathbf{q}^7 \mathbf{t}}
     (\mathbf{a}^2 \mathbf{t}+1) (\mathbf{a}^2
   \mathbf{q}^8 \mathbf{t}^5 (\mathbf{a}^2
   \mathbf{t}+1)+\mathbf{q}^6 \mathbf{t}^2 (\mathbf{a}^6
   \mathbf{t}^5+\mathbf{a}^4 \mathbf{t}^4+\mathbf{a}^2
   \mathbf{t}+1)+\mathbf{a}^2 \mathbf{q}^4 \mathbf{t}^3
   (\mathbf{a}^2
   (\mathbf{t}^3+\mathbf{t})+2)+\mathbf{q}^2
   (\mathbf{a}^6 \mathbf{t}^7+\mathbf{a}^4 \mathbf{t}^4+\mathbf{a}^2
   \mathbf{t}^3+1)+\mathbf{a}^4 \mathbf{t}^4+\mathbf{a}^2
   \mathbf{t})~.$


Let us conclude this section by mentioning the implication of the symmetry \eqref{q-t} of the free energy in the resolved conifold. As the right hand sides of \eqref{refined-GT1} and \eqref{refined-GT2} are interchanged by $(q,t)\leftrightarrow (t^{-1},q^{-1})$, so are the left hand sides. Using the property of Macdonald functions $P_\lambda(-x;t,q)=(-1)^{|\lambda|}P_\lambda(x;t^{-1},q^{-1})$,  this implies
$$
g_\lambda(q,t)~\overline \rCS_\lambda(T_{m,n};a,q,t)=(-1)^{|\lambda|} ~\overline \rCS_{\lambda^T}(T_{m,n};a,t^{-1},q^{-1})~,
$$
which is the \emph{unreduced} version of the mirror/transposition symmetry \eqref{mirror-sym} of the refined Chern-Simons invariants.  This explanation was first provided in \cite[\S3.1]{Aganagic:2012hs}.

\section{Large $N$ duality for torus links}\label{sec:links}
In this section, we generalize refined large $N$ duality to torus links with $L$ components. For each component of a torus link, we introduce $M_i$ M5'-branes supported on the conormal bundle $\ccL_i$ of the component in the deformed conifold $T^*S^3$. These M5'-branes still remain after the geometric transition. Since a half of supersymmetry is preserved if the supports of M5'-branes are special Lagrangian submanifolds in a Calabi-Yau as explained in \S\ref{sec:open}, it is straightforward to extend the analysis in the previous sections to torus links. To avoid repetitious explanation, we shall present only essential results in this section.

For a torus link  $T_{m,n}$ with $\textrm{gcd}(m,n)=L$, we need to introduce the fugacities $x_i$ ($i=1,\cdots,L$) that parametrize the Cartan subgroup of the $\U(M_i)$-valued moduli of $\ccL_i$  both on the deformed conifold and on the resolved conifold. In the resolved conifold, we consider a holomorphic curve $\Sigma_{g,h}$ with $h=\sum_{i=1}^Lh_i$ boundaries where $h_i$ boundaries end on $\ccL_i$. Therefore, we have to project the space $H^*((S^1)^{h-1})\otimes H^*(\cM_{g,h,\beta})$ on the invariant subspace of the relevant symmetry $\frakS_{h_1}\times \cdots \times \frakS_{h_L}$. Eventually, the BPS spectrum for a torus link analogous to \eqref{knot-BPS} is
\begin{multline}\nonumber
{\rm Inv}\Bigl( H^*((S^1)^{h-1})
\otimes H^*({\cal M}_{g,h,\beta})\Bigr)_{\cH(\beta,\mu_1,\cdots,\mu_L)}=\\
\bigoplus_{\{\sigma_i\}\,\{\rho_i\}} C_{\mu_1\,\sigma_1  \rho_1} \cdots C_{\mu_L\,\sigma_L  \rho_L}
{\bf S}_{\sigma_1,\cdots,\sigma_L}(H^*((S^1)^{h-1}))
\otimes {\bf S}_{\rho_1,\cdots,\rho_L}(H^*({\cal M}_{g,h,\beta}))~.
\end{multline}
For the $q$-branes, the fermion zero modes $\psi_{1}^{(i)},\cdots,\psi_{h_i}^{(i)}$ coming from the boundaries ending on $\ccL_i$ contribute to the refined index by $(t^{\frac12}-t^{-\frac12})B_{\sigma_i}(t)$. The factor $(t^{\frac12}-t^{-\frac12})$, which is absent in the case of a torus knot, stems from the fact that we do not impose the linear constraint $\sum_{j=1}^{h_i}\psi_{j}^{(i)}=0$ on each boundary. Hence, we have
\be\label{boundary-link}
{\rm Tr}_{{\bf S}_{\sigma_1,\cdots,\sigma_L}(H^*((S^1)^{h-1}))}\, (-1)^{F}\, q^{J_1-S_R}\, t^{S_R-{J}_2}= (t^{\frac12}-t^{-\frac12})^{L-1}\prod_{i=1}^LB_{\sigma_i}(t)
\ee
where the linear constraint  $\sum_{i=1}^L\sum_{j=1}^{h_i}\psi_{j}^{(i)}=0$ on the total boundary fermion zero modes yields  the factor $(t^{\frac12}-t^{-\frac12})^{-1}$.

As in the case of torus knots, we conjecture that \emph{the extra $\U(1)_R$ global  symmetry $S_R$ acts trivially on the BPS states ${\bf S}_{\rho_1,\cdots,\rho_L}(H^*({\cal M}_{g,h,\beta}))$}. Thus, we decompose the BPS states
 $$
 {\bf S}_{\rho_1,\cdots,\rho_L}(H^*({\cal M}_{g,h,\beta}))\cong \bigoplus _{J_r}A_{\rho_1,\cdots,\rho_L, g,\beta,J_r}
 $$
with respect to only $J_r$ charges but not $S_R$ charges and define
$$
\wh N_{\rho_1,\cdots,\rho_L, g,\beta,J_r} :=\dim A_{\rho_1,\cdots,\rho_L, g,\beta,J_r} ~.
$$

The free energy of the $\bar t$-branes can be obtained from that of the $q$-branes by exchanging $(q,t)\leftrightarrow (t^{-1},q^{-1})$. Therefore, we can write them in the forms
\begin{align}
F^q_{\textrm{ref}}&=\sum_{d>0} \sum_{\{\mu_i\}}\frac{1}{d}\frac{ (t^{\frac{d}{2}}-t^{-\frac{d}{2}})^{L-1}}{q^{\frac{d}{2}}-q^{-\frac{d}{2}}}  f^{q}_{\mu_1,\cdots,\mu_L}(T_{m,n};a^d,q^d,t^d) \prod_{i=1}^L  s_{\mu_i}(x_i^d) ~,\cr
F^{\bar t}_{\textrm{ref}}&= \sum_{d>0} \sum_{\{\mu_i\}}\frac{1}{d}   \frac{(q^{-\frac{d}{2}}-q^{\frac{d}{2}})^{L-1}  }{t^{-\frac{d}{2}}-t^{\frac{d}{2}}}f^{\bar t}_{\mu_1,\cdots,\mu_L}(T_{m,n};a^d,q^d,t^d)\prod_{i=1}^L  s_{\mu_i}(x_i^d) ~,\nonumber
\end{align}
where the refined reformulated invariants take the forms
\begin{align}\label{f-to-hatf-link}
f^{q}_{\mu_1,\cdots,\mu_L}(T_{m,n};a,q,t)&= \sum_{\rho_1,\cdots,\rho_L}M_{\mu_1\rho_1}(t) \cdots M_{\mu_L\rho_L}(t) \wh f_{\rho_1,\cdots,\rho_L}(T_{m,n};a,q,t)~,\cr
f^{\bar t}_{\mu_1,\cdots,\mu_L}(T_{m,n};a,q,t)&=\sum_{\rho_1,\cdots,\rho_L} M_{\mu_1\rho_1}(q^{-1}) \cdots M_{\mu_L\rho_L}(q^{-1}) \wh f_{\rho_1,\cdots,\rho_L}(T_{m,n};a,q,t)~,
 \end{align}
and  $\wh f_{\rho_1,\cdots,\rho_L}(T_{m,n})$ are of the form
  \be\label{hat-f-link}
\wh f_{\rho_1,\cdots,\rho_L}(T_{m,n};a,q,t)=\sum_{\textrm{charges}}  (-1)^{2J_r}{\widehat N}_{\rho_1,\cdots,\rho_L, g,\beta,J_r} (T_{m,n}) (q^{\frac12}-q^{-\frac12})^{g}(t^{-\frac12}-t^{\frac12})^{g} \left(\frac{q}{t} \right)^{J_r-\frac\beta2}a^{\beta}~~
 \ee
 with \emph{non-negative} integers ${\widehat N}_{\rho_1,\cdots,\rho_L, g,\beta,J_r} (T_{m,n})\in\bZ_{\ge0}$. Here we factor out  $(t^{\frac12}-t^{-\frac12})^{L-1}$ in \eqref{boundary-link} from the definition of refined reformulated invariants since it depends only on the number $L$ of link components.

As a result, the large $N$ duality of refined Chern-Simons theory with a torus link $T_{m,n}$ with $L$ components can be summarized as
\begin{multline}\label{Large-N-link-1}
\sum_{\lambda_i} \overline{\textrm{rCS}}_{\lambda_1,\cdots,\lambda_L}(T_{m,n};a,q,t) \prod_{i=1}^L g_{\lambda_i}(q,t) P_{\lambda_i}(x_i;q,t)=    \\
\exp\left(  \sum_{d>0}\sum_{\{\mu_i\}}\frac{1}{d}\frac{ (t^{\frac{d}{2}}-t^{-\frac{d}{2}})^{L-1}}{q^{\frac{d}{2}}-q^{-\frac{d}{2}}}  f^{q}_{\mu_1,\cdots,\mu_L}(T_{m,n};a^d,q^d,t^d) \prod_{i=1}^L  s_{\mu_i}(x_i^d)  \right)~,
\end{multline}
\vspace{-.5cm}
\begin{multline}\label{Large-N-link-2}
\sum_{\lambda_i} \overline{\textrm{rCS}}_{\lambda_1,\cdots,\lambda_L}(T_{m,n};a,q,t) \prod_{i=1}^L  P_{\lambda_i^T}(-x_i;t,q)=\\ \exp\left( \sum_{d>0} \sum_{\{\mu_i\}}\frac{1}{d}   \frac{(q^{-\frac{d}{2}}-q^{\frac{d}{2}})^{L-1}  }{t^{-\frac{d}{2}}-t^{\frac{d}{2}}}f^{\bar t}_{\mu_1,\cdots,\mu_L}(T_{m,n};a^d,q^d,t^d)\prod_{i=1}^L  s_{\mu_i}(x_i^d)  \right)~,
\end{multline}
where the reformulated invariants are of the form \eqref{f-to-hatf-link} with \eqref{hat-f-link}. These identities enable us to express the reformulated invariants in terms of refined Chern-Simons invariants of a torus link, which are presented in \eqref{fq-general} and \eqref{ft-general}. Thus, the large $N$ duality provides a rather non-trivial connection of refined Chern-Simons invariants $\overline \rCS_{\lambda_1,\cdots,\lambda_L}(T_{m,n};a,q,t) $ of a torus link to enumerative invariants  ${\widehat N}_{\rho_1,\cdots,\rho_L, g,\beta,J_r} (T_{m,n})\in\bZ_{\ge0}$ in the resolved conifold.

In Appendix \ref{sec:app-c}, we present some examples of $\wh  N_{\rho_1,\rho_2,g,\beta,J_r}$ for the Hopf link $T_{2,2}$ and the $T_{2,4}$ link by using the results in \cite{DuninBarkowski:2011yx,Gukov:2015gmm}. As in the case of torus knots, one can verify that $2J_r$ charges of non-trivial ${\widehat N}_{\rho_1,\rho_2, g,\beta,J_r}$ are either all even or all odd with $\rho_1,\rho_2,g,\beta$ fixed. Therefore, we conjecture that this is true for any torus link with $L$ components so that the LMOV invariant is
$$
{\widehat N}_{\rho_1,\cdots,\rho_L, g,\beta} (T_{m,n})=\pm\sum_{J_r\in \frac12\bZ}{\widehat N}_{\rho_1,\cdots,\rho_L, g,\beta,J_r} (T_{m,n})~.
$$

\section{Extension to non-torus knots}\label{sec:non-torus}

Let us consider an extension of the formulation above to non-torus knots. In the case of non-torus knots, one option is to consider a generating function of Poincar\'e polynomials of colored HOMFLY-PT homology, which was first examined in \cite{Garoufalidis:2015ewa} in the case of symmetric representations. Recently, the HOMFLY-PT homology colored by arbitrary representations has been defined in \cite{cautis2016remarks} although it is formidable to carry out computation of homology via the definition. In addition, various structural properties of the HOMFLY-PT homology are conjectured \cite{Gukov:2011ry,Gorsky:2013jxa,Gukov:2015gmm,wedrich2016exponential} when colors are specified by rectangular Young diagrams.  Using these properties, conjectural formulas for Poincar\'e polynomials of colored HOMFLY-PT homology have been obtained for a certain class of non-torus knots \cite[references therein]{Gukov:2015gmm}. In \cite{Gorsky:2013jxa,Gukov:2015gmm},  two homological gradings called $\mathbf{t_r}$- and $\mathbf{t_c}$-gradings have been introduced. In the case of torus knots, after the change of variables \eqref{cov}, refined Chern-Simons invariants expressed in terms of the $(\mathbf{a},\mathbf{q},\mathbf{t})$ variables yields $\mathbf{t_c}$-gradings.

As we have seen in the previous sections, the variables $(a,q,t)$ for the equivariant parameters \eqref{omega} are suitable for the formulations of large $N$ duality in refined topological string theory. Hence, for the straightforward extension of refined large $N$ duality to non-torus knots, we consider generating functions of the refined version of HOMFLY-PT polynomials, which we denote by $\overline{\mathscr{P}}_\lambda(K;a,q,t)$. In the case of rectangular Young diagrams, they can be obtained by re-writing Poincar\'e polynomials of colored HOMFLY-PT homology with $\mathbf{t_c}$-grading in term of the $(a,q,t)$ variables by using the change of variables \eqref{cov}. Then, the natural extension of refined large $N$ duality \eqref{refined-GT1} and \eqref{refined-GT2} to non-torus knots is
\begin{align}
\sum_{\lambda} \overline{\mathscr{P}}_\lambda(K;a,q,t)\; g_\lambda(q,t)P_\lambda(x;q,t) &= \exp\left( \sum_{d=1}^\infty \sum_\mu\frac{1}{d}\frac{f^{q}_\mu(K;a^d,q^d,t^d)}{q^{\frac{d}{2}}-q^{-\frac{d}{2}}}    s_\mu(x^d)  \right)~,\label{refined-GT1-nontorus}\\
\sum_{\lambda} \overline{\mathscr{P}}_\lambda(K;a,q,t)~   P_{\lambda^T}(-x;t,q)  &=\exp\left( \sum_{d=1}^\infty \sum_\mu\frac{1}{d}  \frac{f^{\bar t}_{\mu}(K;a^d,q^d,t^d)}{t^{-\frac{d}{2}}-t^{\frac{d}{2}}}    s_{\mu}(x^d)  \right)~.\label{refined-GT2-nontorus}
\end{align}

Recently, closed form expressions of Poincar\'e polynomials of HOMTLY-PT homology colored by (anti-)symmetric representations have been obtained for the $(2s-1,1,2t-1)$-pretzel knots \cite[\S5.3]{Gukov:2015gmm} as well as the knots $\mathbf{6_2}$ and $\mathbf{6_3}$ \cite[\S2.3]{Nawata:2015wya}. Using these data, one can compute reformulated invariants $f^{q}_\mu(K;a,q,t)$ and  $f^{\bar t}_{\mu}(K;a,q,t)$ up to two boxes. Remarkably, it turns out that reformulated invariants of these knots can be brought into the form \eqref{hat-f-2} with \eqref{f-to-hatf-2}.  Some of the resulting BPS degeneracies are tabulated in Appendix \ref{sec:app-c} and more data are included in the \texttt{Mathematica} file.

As an example, let us look at the uncolored BPS degeneracies of the figure-eight presented in Table \ref{table:fig8-uncolor}.
Unlike the case \eqref{resolution} of torus knots, there are (boson-fermion) cancellations by sign in the unrefined limit although they reduce to the corresponding LMOV invariants \cite[Figure 7]{Labastida:2000yw}.  It was observed in \cite{Labastida:2000yw,Labastida:2001ts} that, for any knot $K$, the LMOV invariants ${\widehat N}_{\rho,g,\beta}(K)$ have the same parity of their $a$-gradings $\beta$, \textit{i.e.} $(-1)^\beta{\widehat N}_{\rho,g,\beta}$ become all non-negative integers for any $\mu,g,\beta$ up to appropriate grading shifts.  However, as we see in this example, there are cancellations  behind for non-torus knots, which becomes manifest only at the refined level. This property can be seen in the other examples of the $(2s-1,1,2t-1)$-pretzel knots as well as the knots $\mathbf{6_2}$ and $\mathbf{6_3}$ up to two boxes.

 To go beyond two boxes, we need a $\yng(2,1)$-colored refined invariant, whose definition is not available yet. Nevertheless, we can seek the $\yng(2,1)$-colored refined invariant of the figure-eight which satisfies the mirror symmetry \eqref{mirror-sym} and the exponential growth property \eqref{EGP}. Since the figure-eight knot is amphichiral (a knot which is the same as  its mirror image), we also impose the condition
\be\label{amphichiral}\overline\scP_{\lambda}(\mathbf{4_1};a,q,t)=\overline\scP_{\lambda}(\mathbf{4_1};a^{-1},q^{-1},t^{-1})~.\ee
Like refined Chern-Simons invariants, we however allow that the refined invariant \eqref{fig8-21} can have both positive and negative coefficients even after the change of variables \eqref{cov} so that it cannot be interpreted as a Poincar\'e polynomial. With this condition, we find the \emph{reduced} $\yng(2,1)$-colored refined invariant of the figure-eight:
\begin{footnotesize}
\begin{align}\label{fig8-21}
&\scP_{\yng(2,1)}(\mathbf{4_1};a,q,t)=\frac{a^3 q^{\frac72}}{t^{\frac72}}+a^2
   \left[-\frac{q^{\frac72}}{t^{\frac32}}-\frac{q^{\frac52}}{t^{\frac32}}-\frac{q^{\frac52}}{t^{\frac52}}+\frac{q
   ^{\frac32}}{t^{\frac32}}-\frac{q^{\frac32}}{t^{\frac52}}-\frac{q^{\frac32}}{t^{\frac72}}-\frac{q^{\frac72}}{
   t^{\frac12}}+\frac{q^3}{t}+\frac{q^2}{t^2}-\frac{q^{\frac12}}{t^{\frac72}}+\frac{q}{t^3}\right]\cr
 &  +a   \left[q^{\frac72} t^{\frac32}+\frac{1}{q^{\frac32} t^{\frac72}}+3 q^{\frac52}
   t^{\frac12}+\frac{q^{\frac52}}{t^{\frac12}}-q^{\frac32} t^{\frac12}+\frac{5 q^{\frac32}}{t^{\frac12}}-q^3 t-3
   q^2+\frac{5 q^{\frac12}}{t^{\frac32}}+\frac{q^{\frac12}}{t^{\frac52}}-\frac{1}{q^{\frac12}
   t^{\frac32}}+\frac{3}{q^{\frac12} t^{\frac52}}-\frac{1}{q t^3}-\frac{4
   q}{t}-\frac{3}{t^2}\right]\cr
&  -2 q^{\frac52} t^{\frac52}+2 q^2 t^2+q^{\frac32}
   \left(t^{\frac12}-5 t^{\frac32}\right)+4 q t+q^{\frac12} \left(t^{\frac32}-8
   t^{\frac12}\right)+7+\frac{1}{q^{\frac12}}\left(\frac{1}{t^{\frac32}}-\frac{8}{t^{\frac12}}\right)+\frac{4}{q t}+\frac1{q^{\frac32}}\left(\frac{1}{t^{\frac12}}-\frac{5}{t^{\frac32}}\right)+\frac{2}{q^2
   t^2}-\frac{2}{q^{\frac52} t^{\frac52}}\cr
   &   +a^{-1}\left[q^{\frac32} t^{\frac72}+\frac{1}{q^{\frac72} t^{\frac32}}+\frac{5
   t^{\frac12}}{q^{\frac32}}+\frac{t^{\frac12}}{q^{\frac52}}-\frac{1}{q^{\frac32}
   t^{\frac12}}+\frac{3}{q^{\frac52} t^{\frac12}}-\frac{1}{q^3 t}-\frac{3}{q^2}+3 q^{\frac12}
   t^{\frac52}+\frac{t^{\frac52}}{q^{\frac12}}-q^{\frac12} t^{\frac32}+\frac{5 t^{\frac32}}{q^{\frac12}}-q
   t^3-\frac{4 t}{q}-3 t^2\right]\cr
 &  +a^{-2}\left[
   -\frac{t^{\frac72}}{q^{\frac32}}-\frac{t^{\frac52}}{q^{\frac32}}-\frac{t^{\frac52}}{q^{\frac52}}+\frac{t^{\frac32}
   }{q^{\frac32}}-\frac{t^{\frac32}}{q^{\frac52}}-\frac{t^{\frac32}}{q^{\frac72}}-\frac{t^{\frac12}}{q^{\frac72}}+\frac{t}{q^3}+\frac{t^2}{q^2}-\frac{t^{\frac72}}{q^{\frac12}}+\frac{t^3}{q}\right]+\frac{t^{\frac72}}{a^3 q^{\frac72}}~.
   \end{align}
\end{footnotesize}
Surprisingly,  the reformulated invariants computed by using this datum also can be written in the form of \eqref{hat-f-2} with \eqref{f-to-hatf-2}, and the corresponding BPS degeneracies are presented in Appendix \ref{sec:app-c}. Because of the properties \eqref{amphichiral} and \eqref{mac-inverse}, the left hand sides of \eqref{refined-GT1-nontorus} and \eqref{refined-GT2-nontorus} stay invariant under the change of variables  $(a,q,t)\leftrightarrow (a^{-1},q^{-1},t^{-1}) $ for an amphichiral knot $K$. Then, the right hand sides are also invariant if
$$
\wh f_\rho(K;a^{-1},q^{-1},t^{-1})  = (-1)^{|\mu|}\; \wh f_{\mu^T}(K;a,q,t)
$$
due to \eqref{B-mirror}, which is equivalent to the condition
$$\wh  N_{\rho,g,\beta,J_r}(K) =\wh  N_{\rho^T,g,-\beta,-J_r}(K)~, $$
to the BPS degeneracies (up to overall sign). This property can be seen for amphichiral knots like the figure-eight and the knot $\mathbf 6_3$ in Appendix \ref{sec:app-c}.

Let us mention the cases in which situations are different. It is known that the  $(2s-1,1,2t-1)$-pretzel knots as well as the knots $\mathbf{6_2}$ and $\mathbf{6_3}$ are homologically-thin and their HOMFLY-PT homology colored by rectangular Young diagrams is subject to the exponential growth property. On the other hand, the \emph{thick} HOMFLY-PT homology has more complicated properties and it is \emph{not} endowed with the exponential growth property. This was shown in \cite[Appendix B]{Gukov:2011ry} by explicitly obtaining the Poincar\'e polynomials of the $\yng(2)$-colored HOMFLY-PT homology with $\mathbf{t_r}$-grading of the knot $\mathbf{9_{42}}$, which is a non-torus homological-thick knot with the fewest crossing.  It is a straightforward exercise to obtain both $\yng(2)$-colored and $\yng(1,1)$-colored HOMFLY-PT homology with $\mathbf{t_c}$-grading for the knot $\mathbf{9_{42}}$ and the corresponding refined invariants.  It turns out that the reformulated invariants of  the knot $\mathbf{9_{42}}$ obtained by these data \emph{cannot} be expressed in the desired form \eqref{hat-f-2} with \eqref{f-to-hatf-2}.

Moreover, the HOMFLY-PT homology of some non-torus links including the Whitehead link has been obtained in \cite{Gukov:2015gmm}. However, the straightforward extension of \eqref{Large-N-link-1} and \eqref{Large-N-link-2} to any non-torus links fails to provide the desired form of the reformulated invariants even in the fundamental representation.

\section{Discussion}\label{sec:discussion}

In this paper, we have formulated large $N$ duality of refined Chern-Simons theory with a torus knot/link. Assuming that the extra $\U(1)_R$ global symmetry acts trivially on the BPS states coming from deformations of M2-branes, this formulation gives a striking relation between refined Chern-Simons invariants of a torus knot/link and graded dimensions of cohomology groups of moduli spaces of M2-M5 bound states in the resolved conifold. Therefore, this leads to the \emph{positivity conjecture} of refined Chern-Simons invariants of a torus knot/link. Conversely, one can obtain complete information about BPS spectra in M-theory on the resolved conifold with M5'-branes supported on $\bR^3\times\ccL_{T_{m,n}}$ by using the geometric transition. It is also worth mentioning that, for M-theory on any non-compact toric Calabi-Yau three-fold with M5-branes, its free energy on the $\Omega$-background takes the forms \eqref{free-q} and \eqref{free-t} if the extra $\U(1)_R$ global symmetry is preserved. It is important to understand in which situations the extra $\U(1)_R$ global symmetry acts on the space of BPS states trivially as in this paper.

As we have seen in \S\ref{sec:non-torus}, the refined large $N$ duality can be extended to a certain class of homologically-thin non-torus knots. However, we checked that this does not work for homologically-thick non-torus knots as well as any non-torus links. These results are still at the level of observation and the underlying structure needs to be investigated.

Giving a mathematical definition of refined LMOV invariants ${\widehat N}_{\rho, g,\beta,J_r} (T_{m,n})$ discussed in this paper is a challenging, but important open problem. Refined GV invariants have been discussed in the literature \cite[references therein]{Choi:2012jz,Chuang:2013wpa,Nekrasov:2014nea,Gu:2017ccq} as refined Pandharipande-Thomas (Donaldson-Thomas) invariants for non-compact toric Calabi-Yau three-folds. However, mathematical understanding of their open analogues are still immature although the Poincar\'e polynomials of uncolored HOMFLY-PT homology of torus knots have been related to motivic Donaldson-Thomas invariants in \cite{Diaconescu:2012dw}.  Actually, upon the reduction on the cigar of the Taub-NUT in \eqref{resolved}, BPS states can be understood as  D6-D4-D2-D0 bound states. Hence, it is an important task to give a mathematical definition of D6-D4-D2-D0 bound states for refined Chern-Simons invariants discussed in this paper. Besides, it would be also intriguing to find a connection to (a certain variant of) ``P=W conjecture'' \cite[references therein]{Chuang:2013wpa,Diaconescu:2017tga}.

Another direction to pursue is to find large $N$ duality of refined Chern-Simons theory with different gauge groups. In fact, $\SO(2N)$ refined Chern-Simons theory has been proposed \cite{Aganagic:2012au}, generalizing Kauffman polynomials. In addition, large $N$ duality for Kauffman polynomials has been put forward by incorporating orientifolds in the resolved conifold \cite{Marino:2009mw}. Consequently, this leads to an integrality conjecture involving both colored Kauffman and HOMFLY-PT polynomials. It is natural to ask whether a positivity property can be seen when the conjecture of \cite{Marino:2009mw} is refined.

\section*{Acknowledgements}
The authors are indebted to P. Su{\l}kowski for collaboration at the initial stage of this project. The motivation of this paper originated from the paper \cite{Garoufalidis:2015ewa}.
They would like to thank H. Awata, M. Dedushenko, S. Gukov, H. Kanno, A. Klemm, M. Mari\~no, A. Morozov, A. Mironov, P. Putrov,  P. Ramadevi, I. Saberi, A. Sleptsov, Zodinmawia for discussion and correspondence.  In particular, they are grateful to S. Shakirov for sharing Maple files of refined Chern-Simons invariants obtained in \cite{Shakirov:2013moa}.

S.N. would like to thank the American Institute of Mathematics for supporting this research with a SQuaRE grant ``New connections between link homologies and physics''. In addition, S.N. is grateful to University of Science and Technology of China,  Yau Mathematical Sciences Center at Tsinghua University and Nagoya University for invitation and hospitality.
The work of S.N. is  supported by  ERC Starting Grant no.~335739 \textit{``Quantum fields and knot homologies''}, the center of excellence grant ``Center for Quantum Geometry of Moduli Space'' from the Danish National Research Foundation (DNRF95),  and  Walter Burke Institute for Theoretical Physics, California Institute of Technology as well as the U.S. Department of Energy, Office of Science, Office of High Energy Physics, under Award Number DE-SC0011632.

\allowdisplaybreaks
\appendix

\section{Symmetric functions}\label{sec:app-a}

\begin{wrapfigure}{R}{0.4\textwidth}
\centering
\begin{tikzpicture}[scale=0.80]
\draw (0,0)--(0,-5)--(1,-5)--(1,-4)--(3,-4)--(3,-2)--(5,-2)--(5,-1)--(6,-1)--(6,0)--(0,0);
\draw (2,-1.25)--(2,-1.75)--(2.5,-1.75)--(2.5,-1.25)--(2,-1.25);
\draw (2.25,-1.5) node {$s$};
\draw [<->,>=stealth] (2.5,-1.5)--(5,-1.5);
\draw [<->,>=stealth] (0,-1.5)--(2,-1.5);
\draw [<->,>=stealth] (2.25,-1.75)--(2.25,-4);
\draw [<->,>=stealth] (2.25,0)--(2.25,-1.25);
\draw (3.8,-1.25) node {$a(s)$};
\draw (1,-1.25) node {$a'(s)$};
\draw (1.9,-2.8) node {$l(s)$};
\draw (2.8,-0.7) node {$l'(s)$};
\end{tikzpicture}
\caption{Arm, leg, co-arm and co-leg}
\label{fig:YT}
\end{wrapfigure}
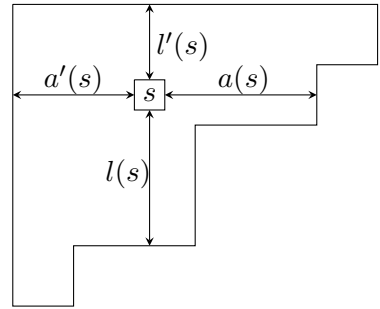

In this appendix, we review basics of symmetric functions relevant to this paper. For more detail, we refer the reader to  \cite{macdonald1998symmetric}.

Let $x=\left(x_1,x_2,\cdots\right)$ be an infinite number of the variables, $\lambda=\left(\lambda_1,\lambda_2,\cdots\right)$ be a Young diagram (i.e. non-negative integers such that $\lambda_i \geq \lambda_{i+1}$ and $|\lambda|=\sum_i\lambda_i<\infty$) and $\vec{k}=(k_1,k_2,\cdots)$ be a vector with a infinite number of entries, almost zero, and whose nonzero entries are positive integers. The  Young diagram $\lambda$ and the vector $\vec{k}$ are in one-to-one correspondence with the relation $k_i=m_i\left(\lambda\right)$, where $m_i\left(\lambda\right)$ is a multiplicity of $i$ in $\lambda$.

First, we define the power-sum symmetric functions by $p_d\left(x\right)=\sum_{i=1}^{\infty}x_i^d$. It is convenient to denote their products by $p_\lambda\left(x\right)=\prod_i p_{\lambda_i}\left(x\right)$ and $p_{\vec{k}}=\prod_i p_{i}^{k_i}$. These are bases of the ring of symmetric functions. Schur and Macdonald functions can be defined by introducing an inner product on the ring of symmetric functions.

\subsection*{Schur functions}

\noindent The Schur functions $s_\lambda\left(x\right)$ are uniquely defined by orthogonality and normalization conditions:
\begin{align}\nonumber
&\langle s_\lambda,s_\mu\rangle=0~, &\mathrm{if} \ \ \lambda\neq \mu~,\\
&s_\lambda\left(x\right)=w_\lambda\left(x\right)+ \sum_{\mu<\lambda}u_{\lambda\mu}w_\mu\left(x\right)~, &u_{\lambda\mu}\in\mathbb{Q}~,\nonumber
\end{align}
where $w_\lambda\left(x\right)$ is the monomial symmetric function, $<$ is dominance partial ordering ($\lambda\geq\mu \Leftrightarrow |\lambda|=|\mu|$ and $\sum_{i=1}^k\lambda_i\geq\sum_{i=1}^k\mu_i$ for all $i$), and
the inner product is defined by
\begin{align}\nonumber
\langle p_\lambda, p_\mu\rangle=\delta_{\lambda\mu}z_\lambda
~, \hspace{20pt}z_\lambda=\prod_{i\geq1}i^{m_i}m_i!~,
\end{align}
where $m_i=m_i\left(\lambda\right)$ is a multiplicity of $i$ in $\lambda$.

The relation between Schur and power sum symmetric functions is known as Frobenius formula:
\begin{align}\label{Schur-Newton}
&s_\lambda\left(x\right)=\sum_{\vec{k}}\frac{\chi_\lambda(C(\vec{k}))}{z_{\vec{k}}}p_{\vec{k}}\left(x\right)~,
&p_{\vec{k}}\left(x\right)=\sum_\lambda\chi_\lambda(C(\vec{k}))s_\lambda\left(x\right)~,
\end{align}
where $\chi_\lambda(C(\vec{k}))$ is the character of the representation $\lambda$ of the permutation group $\mathfrak{S}_h$ evaluated at the conjugacy class $C(\vec{k})$ where $h=\sum_j j k_j$.
The Frobenius formula is used for the computation of Clebsch-Gordon coefficients \eqref{CG} of  the permutation group $\mathfrak{S}_h$.

\subsection*{Macdonald functions}

The Macdonald functions $P_\lambda\left(x;q,t\right)$ are  uniquely defined by orthogonality and normalization conditions:
\begin{equation}\nonumber
\begin{array}{ll}
\langle P_\lambda,P_\mu\rangle_{q,t}=0~, &\qquad\mathrm{if} \ \ \lambda\neq \mu,\\
P_\lambda\left(x;q,t\right)=w_\lambda\left(x\right)+ \sum_{\mu<\lambda}u_{\lambda\mu}\left(q,t\right)w_\mu\left(x\right)~, &\qquad u_{\lambda\mu}\left(q,t\right)\in\mathbb{Q}\left(q,t\right)~,
\end{array}
\end{equation}
where the inner product is defined by
\begin{align}\nonumber
\langle p_\lambda, p_\mu\rangle_{q,t}=\delta_{\lambda\mu}z_\lambda\prod_{i\geq1}\frac{1-q^{\lambda_i}}{1-t^{\lambda_i}}~, \hspace{20pt}z_\lambda=\prod_{i\geq1}i^{m_i}m_i!~.
\end{align}
At the $q=t$ specialization, the Macdonald functions reduce to the Schur functions. From the definition one can show
$$
\frac{(q/t)^{|\lambda|}}{g_\lambda(q,t)}:=\langle P_\lambda, P_\lambda\rangle_{q,t}=\prod_{s\in\lambda}\frac{1-q^{a(s)+1}t^{l(s)}}{1-q^{a(s)}t^{l(s)+1}}~,
$$
where an arm length $a\left(s\right)=\lambda_i-j$ and a leg length $l\left(s\right)=\lambda^T_j-i$ for each box $s=\left(i,j\right)$ in $\lambda$ are depicted in Figure \ref{fig:YT}.

We denote by $\frakX_\lambda(\vec{k};q,t)$ coefficients in the expansion of a Macdonald function $P_\lambda\left(x;q,t\right)$ with respect to $p_{\vec{k}}\left(x\right)$:
$$
P_\lambda\left(x;q,t\right)=\sum_{\vec{k}}\frakX_\lambda(\vec{k};q,t)p_{\vec{k}}\left(x\right)~.
$$
Since the rational functions $\frakX_\lambda(\vec{k};q,t)$ are invariant under the exchange $(q,t)\leftrightarrow(q^{-1},t^{-1})$, the Macdonald functions have the following property
\be\label{mac-inverse}  P_\lambda\left(x;q,t\right)=P_\lambda\left(x;q^{-1},t^{-1}\right)~.\ee
Note that at the $q=t$ specialization
$$
\frakX_\lambda(\vec{k};q,q)=\frac{\chi_\lambda(C(\vec{k}))}{z_{\vec{k}}}~.
$$
In the following, we list some Macdonald functions expressed in terms of the power-sum functions:
\begin{align}\nonumber
P_{\yng(1)} = &p_1\cr
P_{\yng(2)} =& \frac{(1-t)(1+q)}{(1-tq)} \frac{p_1^2}{2} + \frac{(1+t)(1-q)}{(1-tq)} \frac{p_2}{2}, \cr
P_{\yng(1,1)}=&\frac{p_1^2}{2} - \frac{p_2}{2}\cr
P_{\yng(3)} =& \frac{(1+q)(1-q^3)(1-t)^2}{(1-q)(1-tq)(1-tq^2)} \frac{p_1^3}{6} + \frac{(1-q)(1-t^2)(1-q^3)}{(1-q)(1-tq)(1-tq^2)} \frac{p_1p_2}{2} \cr
&+
\frac{(1-q)(1-q^2)(1-t^3)}{(1-t)(1-tq)(1-tq^2)} \frac{p_3}{3}\cr
P_{\yng(2,1)} =& \frac{(1-t)(2qt + q + t + 2)}{1-qt^2} \frac{p_1^3}{6} + \frac{(1+t)(t-q)}{1-qt^2} \frac{p_1p_2}{2} -
\frac{(1-q)(1-t^3)}{(1-t)(1-qt^2)} \frac{p_3}{3}\cr
P_{\yng(1,1,1)} =& \frac{p_1^3}{6} - \frac{p_2p_1}{2} + \frac{p_3}{3}.
\end{align}
For instance, we can read off $\frakX_{\yng(2,1)}({\vec{k}=(3,0,0)};q,t)=\frac{(1-t)(2qt + q + t + 2)}{6({1-qt^2})}$ because of $p_{\vec{k}={(3,0,0)}}=p_1^3$.

\subsection*{Cauchy formulas}
The Cauchy formulas play a very important role in this paper. The Cauchy formulas for Schur functions read off:
\begin{align}\nonumber
\sum_\lambda s_\lambda(x)s_\lambda(y)=\exp\left(\sum_{d>0}\frac{1}{d}p_d(x)p_d(y)\right)~,\quad
\sum_\lambda s_\lambda(x)s_{\lambda^T}(y)=\exp\left(\sum_{d>0}\frac{(-1)^{d-1}}{d}p_d(x)p_d(y)\right)~.
\end{align}
The analogues of Macdonald functions are
\begin{align}\label{Cauchy-mac}
\sum_\lambda g_\lambda(q,t) P_\lambda(x;q,t)P_\lambda(y;q,t)&=\exp\left(\sum_{d>0}\frac{1}{d}\frac{t^{\frac{d}{2}}-t^{-\frac{d}{2}}}{q^{\frac{d}{2}}-q^{-\frac{d}{2}}}p_d(x)p_d(y)\right)~,\cr
\sum_\lambda P_\lambda(x;q,t)P_{\lambda^T}(y;t,q)&=\exp\left(\sum_{d>0}\frac{(-1)^{d-1}}{d}p_d(x)p_d(y)\right)~.
\end{align}

\section{Explicit formulas of refined reformulated invariants}\label{sec:app-b}

In this appendix, we will derive explicit formulas for refined reformulated invariants of a torus link $T_{m,n}$ with $L$ components in terms of its refined Chern-Simons invariants from  \eqref{Large-N-link-1} and  \eqref{Large-N-link-2} by following  \cite{Labastida:2001ts}. To this end, we define the plethystic exponential and its inverse
$$
 \mathrm{Exp}(F):=\exp\left(\sum_{d=1}^\infty \frac{\psi_d}{d}\right)\circ F~, \qquad
 \mathrm{Log}(F):=\sum_{d=1}^\infty\frac{\mu(d)}{d}\log(\psi_d\circ F)~,
$$
where an operator $\psi_d$ is defined by $\psi_d\circ
F(a,q,t;x):=F(a^d,q^d,t^d;x^d)$ and $\mu(d)$ is the M\"obius
function. If one sets
$$
F:=
 \frac{(t^{\frac{1}{2}}-t^{-\frac{1}{2}})^{L-1}}
 {q^{\frac{1}{2}}-q^{-\frac{1}{2}}} \sum_{\{\mu_i\}}f^{q}_{\mu_1,\cdots,\mu_L}\left(T_{m,n};a,q,t \right)
 \prod_{i=1}^L s_{\mu_i}(x_i)~,
$$
then the right hand side of  \eqref{Large-N-link-1} can be written as $\mathrm{Exp}(F)$. Thus, one can manipulate the identity  \eqref{Large-N-link-1} as
\begin{align}
F&=
 \mathrm{Log}\left(
 \sum_{\{\lambda_i\}}\overline{\rCS}_{\lambda_1\cdots\lambda_L}(T_{m,n};a,q,t)
 \prod_{i=1}^Lg_{\lambda_i}(q,t)P_{\lambda_i}(x_i;q,t)
 \right)\cr
 &=\sum_{d=1}^\infty\frac{\mu(d)}{d}\log\left(
 \sum_{\{\lambda_i\}}\overline{\rCS}_{\lambda_1\cdots\lambda_L}^{(d)}
 \prod_{i=1}^Lg_{\lambda_i}(q^d,t^d)P_{\lambda_i}(x_i^d;q^d,t^d)
 \right)\cr
 &=\sum_{d=1}^\infty\frac{\mu(d)}{d}
 \sum_{m=1}^\infty
 \frac{(-1)^{m-1}}{m}
 \prod_{\alpha=1}^m
 \sum_{\{\lambda_i^{(\alpha)}\}}\overline{\rCS}_{\lambda_1^{(\alpha)}\cdots\lambda_L^{(\alpha)}}^{(d)}
 \prod_{i=1}^Lg_{\lambda_i^{(\alpha)}}(q^d,t^d)P_{\lambda_i^{(\alpha)}}(x_i^d;q^d,t^d)\cr
  &=\sum_{d=1}^\infty\frac{\mu(d)}{d}
 \sum_{m=1}^\infty
 \frac{(-1)^{m-1}}{m}
 \prod_{\alpha=1}^m
 \sum_{\{\lambda_i^{(\alpha)}\}}
 \overline{\rCS}_{\lambda_1^{(\alpha)}\cdots\lambda_L^{(\alpha)}}^{(d)}
 \prod_{i=1}^Lg_{\lambda_i^{(\alpha)}}(q^d,t^d)
 \sum_{\vec{k}^{(\alpha)}_i}
\mathfrak{X}_{\lambda_i^{(\alpha)}}(\vec{k}^{(\alpha)}_i;q^d,t^d)
 p_{\vec{k}^{(\alpha)}_i}(x^d_i)~,\nonumber
\end{align}
where $\overline \rCS^{(d)}_\lambda=\overline \rCS_\lambda(T_{m,n};a^d,q^d,t^d)$ and other notations are given in Appendix \ref{sec:app-a}.
To compare with the coefficient of $\prod_{i=1}^{L}s_{\mu_i}(x_i)$, we introduce $\vec{k}_d$ for $\vec{k}=(k_1,k_2,\cdots)$ as
$(\vec{k}_d)_{di}=(\vec{k})_i$, \textit{i.e.}
$$
 \vec{k}_d=(0,\cdots,0,k_1,0,\cdots,0,k_2,0,\cdots)~,
$$
where $k_1$ is $d$-th entry, $k_2$ is $2d$-th entry and so on.
Then, the properties $p_{\vec{k}}(x^d)=p_{\vec{k}_d}(x)$ and
$p_{\vec{k}}(x)p_{\vec{k'}}(x)=p_{\vec{k}+\vec{k'}}(x)$ of the power sum functions tell us
$$
 \prod_{\alpha=1}^m  p_{\vec{k}^{(\alpha)}_i}(x^d_i)=
 p_{\sum_{\alpha=1}^{m}(\vec{k}^{(\alpha)}_i)_d}(x_i)
 =
 \sum_{\{\mu_i\}}
 \chi_{\mu_i} (C(\sum_{\alpha=1}^{m}(\vec{k}^{(\alpha)}_i)_d))s_{\mu_i}(x_i)~.
$$
Using these results, we obtain the explicit formula for $f^q$ in terms of refined Chern-Simons invariants:
\begin{align}\label{fq-general}
& \frac{(t^{\frac{1}{2}}-t^{-\frac{1}{2}})^{L-1}}
 {q^{\frac{1}{2}}-q^{-\frac{1}{2}}}   f^{q}_{\mu_1,\cdots,\mu_L}\left(T_{m,n};a,q,t \right)=
\\& \sum_{d,m=1}^\infty(-1)^{m-1}\frac{\mu(d)}{d\cdot m}
 \sum_{\{\vec{k}^{(\alpha)}_i\}}\sum_{\{\lambda_i^{(\alpha)}\}}
 \prod_{i=1}^L
 \chi_{\mu_i} (C(\sum_{\alpha=1}^{m}(\vec{k}^{(\alpha)}_i)_d))
 \prod_{\alpha=1}^m
 g_{\lambda_i^{(\alpha)}}(q^d,t^d)
\frakX_{\lambda_i^{(\alpha)}}(\vec{k}^{(\alpha)}_i;q^d,t^d)
 \overline{\rCS}_{\lambda_1^{(\alpha)}\cdots\lambda_L^{(\alpha)}}^{(d)}~.\nonumber
\end{align}
Similarly, we can obtain the explicit formula for $f^{\bar{t}}$:
\begin{align}\label{ft-general}
&  \frac{(q^{-\frac{d}{2}}-q^{\frac{d}{2}})^{L-1}  }{t^{-\frac{d}{2}}-t^{\frac{d}{2}}}  f^{\bar{t}}_{\mu_1,\cdots,\mu_L}\left(T_{m,n};a,q,t \right)=\\
& \sum_{d,m=1}^\infty(-1)^{m-1}\frac{\mu(d)}{d\cdot m}
 \sum_{\{\vec{k}^{(\alpha)}_i\}}\sum_{\{\lambda_i^{(\alpha)}\}}
 \prod_{i=1}^L
 \chi_{\mu_i^T} (C(\sum_{\alpha=1}^{m}(\vec{k}^{(\alpha)}_i)_d))
 \prod_{\alpha=1}^m
  (-1)^{|\lambda_i^{(\alpha)}|}
 \frakX_{(\lambda_i^{(\alpha)})^T}(\vec{k}^{(\alpha)}_i;t^d,q^d)
 \overline{\rCS}_{\lambda_1^{(\alpha)}\cdots\lambda_L^{(\alpha)}}^{(d)}~.\nonumber
\end{align}

In the following, we provide reformulated invariants of a torus knot  colored by Young diagrams with a few boxes for the $q$-branes from \eqref{fq-general}:
\begin{align}\nonumber
\frac{f^{q}_{\yng(1)}}{t^{\frac12}-t^{-\frac12}}=&\overline \rCS_{\yng(1)},\\
\frac{t^{\frac12}}{q^{\frac12}} \frac{f^{q}_{\yng(2)}}{t^{\frac12}-t^{-\frac12}}=&
 \frac{q t-1}{q^2-1}\overline  \rCS_{\yng(2)}
-\frac{t-1}{2 (q-1)}(\overline\rCS_{\yng(1)})^2
-\frac{t+1}{2 (q+1)}\overline \rCS_{\yng(1)}^{(2)}~,\nonumber\\
\frac{t^{\frac12}}{q^{\frac12}}\frac{f^{q}_{\yng(1,1)}}{t^{\frac12}-t^{-\frac12}}=&
\frac{t-q}{q^2-1}\overline  \rCS_{\yng(2)}
+\frac{t^2-1}{q t-1}\overline \rCS_{\yng(1,1)}
-\frac{t-1}{2 (q-1)}(\overline\rCS_{\yng(1)})^2
 +\frac{t+1}{2(q+1)}\overline \rCS_{\yng(1)}^{(2)}~,\nonumber\\
\frac{t}{q}\frac{f^{q}_{\yng(3)}}{t^{\frac12}-t^{-\frac12}}=&
\frac{(q t-1) \left(q^2 t-1\right)}{(q^2-1)
   \left(q^3-1\right)}\overline \rCS_{\yng(3)}
-\frac{(t-1) (q t-1)}{(q-1)^2
   (q+1)}\overline \rCS_{\yng(2)}
\overline \rCS_{\yng(1)}\nonumber\\
&+\frac{(t-1)^2}{3 (q-1)^2}(\overline \rCS_{\yng(1)})^3
-\frac{t^2+t+1}{3 \left(q^2+q+1\right)}\overline
 \rCS_{\yng(1)}^{(3)}~,
\nonumber\\
\frac{t}{q}\frac{f^{q}_{\yng(2,1)}}{t^{\frac12}-t^{-\frac12}}=&
-\frac{(q-t) (q t-1)}{(q-1)(q^3-1)}\overline \rCS_{\yng(3)}
+\frac{(t-1) \left(q t^2-1\right)}{(q-1) \left(q^2 t-1\right)}\overline \rCS_{\yng(2,1)}\nonumber\\
&-\left[\frac{(t-1)^2
   }{(q-1)^2}\overline \rCS_{\yng(2)}
+\frac{(t-1)^2 (t+1) }{(q-1) (q
   t-1)}\overline \rCS_{\yng(1,1)}
\right]\overline \rCS_{\yng(1)}\cr
&
 +\frac{2
   (t-1)^2}{3 (q-1)^2}(\overline \rCS_{\yng(1)})^3
+\frac{t^2+t+1}{3 \left(q^2+q+1\right)}\overline \rCS_{\yng(1)}^{(3)}~,
\nonumber\\
\frac{t}{q}\frac{f^{q}_{\yng(1,1,1)}}{t^{\frac12}-t^{-\frac12}}=&\frac{(q-t) \left(q^2-t\right)}{(q^2-1) (q^3-1)}\overline \rCS_{\yng(3)}
-\frac{(t^2-1) (q-t)}{(q-1) \left(q^2 t-1\right)}\overline \rCS_{\yng(2,1)}
 +\frac{(t^2-1) (t^3-1) }{(q
   t-1) \left(q t^2-1\right)}\overline \rCS_{\yng(1,1,1)}\cr
   &
+ \left[  \frac{(t-1) (q-t) }{(q-1)^2
   (q+1)}\overline \rCS_{\yng(2)}
 -\frac{(t-1)^2 (t+1) }{(q-1) (q t-1)}\overline \rCS_{\yng(1,1)}\right]
\overline \rCS_{\yng(1)}\nonumber\\
&
 +\frac{(t-1)^2}{3 (q-1)^2}(\overline \rCS_{\yng(1)})^3
-\frac{t^2+t+1}{3 \left(q^2+q+1\right)}\overline \rCS_{\yng(1)}^{(3)}~.\nonumber
\end{align}
In addition, we present reformulated invariants of a torus knot colored by Young diagrams with a few boxes for the $\bar t$-branes  from \eqref{ft-general}:
 \begin{align}\nonumber
 \frac{f^{\bar t}_{\yng(1)}}{t^{\frac12}-t^{-\frac12}}=&\overline \rCS_{\yng(1)},\\
\frac{- f^{\bar t}_{\yng(2)}}{t^{\frac12}-t^{-\frac12}}=&
\overline{\rCS}_{\yng(1,1)}+
   \frac{1}{2}\overline{\rCS}_{\yng(1)}^{(2)}
   -\frac{1}{2}(\overline{\rCS}_{\yng(1)})^2~,\nonumber\\
\frac{- f^{\bar t}_{\yng(1,1)}}{t^{\frac12}-t^{-\frac12}}=&
\overline{\rCS}_{\yng(2)}
+\frac{q-t}{q t-1}\overline{\rCS}_{\yng(1,1)}
-\frac{1}{2}\overline{\rCS}_{\yng(1)}^2
-\frac{1}{2}\overline{\rCS}_{\yng(1)}^{(2)}
,\nonumber\\
  \frac{f^{\bar t}_{\yng(3)}}{t^{\frac12}-t^{-\frac12}}=&
  \overline{\rCS}_{\yng(1,1,1)}-\overline{\rCS}_{\yng(1)}\overline{\rCS}_{\yng(1,1)}+\frac{1}{3}(\overline{\rCS}_{\yng(1)})^3-\frac{1}{3}\overline{\rCS}_{\yng(1)}^{(3)}~,\nonumber\\
 \frac{f^{\bar t}_{\yng(2,1)}}{t^{\frac12}-t^{-\frac12}}=&
\overline{\rCS}_{\yng(2,1)}
+\frac{(t+1) (q-t)}{q t^2-1}\overline{\rCS}_{\yng(1,1,1)}
-\left[\overline{\rCS}_{\yng(2)} +\frac{(q-1) (t+1)}{q t-1}\overline{\rCS}_{\yng(1,1)}\right]\overline{\rCS}_{\yng(1)}
\nonumber\\&
+\frac{2}{3} (\overline{\rCS}_{\yng(1)})^3+\frac{1}{3} \overline{\rCS}_{\yng(1)}^{(3)},\nonumber\\
\frac{f^{\bar t}_{\yng(1,1,1)}}{t^{\frac12}-t^{-\frac12}}=&\overline{\rCS}_{\yng(3)}
+\frac{(q+1) (q-t)}{q^2 t-1}\overline{\rCS}_{\yng(2,1)}
+\frac{(q-t) \left(q-t^2\right)}{(q t-1) \left(q t^2-1\right)}\overline{\rCS}_{\yng(1,1,1)}\nonumber\\
&+\left[-\overline{\rCS}_{\yng(2)} +\frac{(q-t)}{q t-1}\overline{\rCS}_{\yng(1,1)}\right]\overline{\rCS}_{\yng(1)}
+\frac{1}{3} (\overline{\rCS}_{\yng(1)})^3-\frac{1}{3} \overline{\rCS}_{\yng(1)}^{(3)}~.\nonumber
 \end{align}

\begin{landscape}
\section{Tables of BPS degeneracies}\label{sec:app-c}
In this appendix, we shall list tables of non-negative integral invariants $\widehat N_{\rho_1,\cdots,\rho_L,g,\beta,J_r}$ of some torus knots/links as well as non-torus knots. The prescription to compute these integral invariants is explained in \S\ref{sec:knots} for torus knots,  \S\ref{sec:links} for torus links, and \S\ref{sec:non-torus} for non-torus knots, respectively. We also note that the \texttt{Mathematica} file includes more data.

One can verify that $2J_r$ charges of non-trivial ${\widehat N}_{\rho, g,\beta,J_r}$ are either all even or all odd with $\rho,g,\beta$ fixed for torus knots. (The situation for torus links is the same.) However, this is not true for ${\widehat N}_{\rho, g,\beta,J_r}$ of non-torus knots. In addition, one can see $\wh  N_{\rho,g,\beta,J_r}(K) =\wh  N_{\rho^T,g,-\beta,-J_r}(K) $ for the amphichiral knot $\mathbf{4_1}$ and $\mathbf{6_3}$. (Strictly speaking, when the number of boxes of a Young diagram $\rho$ is odd, one needs to shift $\beta$ and $2J_r$ by $\frac12$ to see this symmetry.)

\subsection*{Torus knots}

  \begin{table}[h]
    \begin{minipage}[c]{7cm}\centering

\begin{tabular}{|c|ccc|ccc|c|}
\hline
 $\beta$& \multicolumn{3}{c|}{1} & \multicolumn{3}{c|}{2} & \multicolumn{1}{c|}{3} \\
 \hline $g\backslash 2J_r$ & 0 & 1 & 2 & 1 & 2 & 3 & 4 \\
 \hline
 0 & 1 & 0 & 1 & 1 & 0 & 2 & 1 \\
 1 & 0 & 1 & 0 & 0 & 1 & 0 & 0 \\
\hline
\end{tabular}
\caption{$\widehat{{N}}_{[1],g,\beta,J_r}(T_{2,3})$ for the trefoil}
    \end{minipage}
    \begin{minipage}[c]{16cm}\centering
\begin{tabular}{|c|ccc|ccccc|ccccccc|ccccc|ccc|}
\hline
 $\beta$ & \multicolumn{3}{c|}{1} & \multicolumn{5}{c|}{2} & \multicolumn{7}{c|}{3} & \multicolumn{5}{c|}{4} & \multicolumn{3}{c|}{5}  \\
 \hline $g\backslash 2J_r$ & 0 & 1 & 2 & 1 & 2 & 3 & 4 & 5 & 2 & 3 & 4 & 5 & 6 & 7 & 8
 & 5 & 6 & 7 & 8 & 9 & 8 & 9 & 10 \\
 \hline
 0 & 1 & 0 & 1 & 2 & 0 & 4 & 0 & 2 & 1 & 0 & 5 & 0 & 5 & 0 & 1 & 2 & 0 & 4 & 0
   & 2 & 1 & 0 & 1 \\
 1 & 0 & 1 & 0 & 0 & 3 & 0 & 3 & 0 & 0 & 2 & 0 & 6 & 0 & 2 & 0 & 0 & 3 & 0 & 3
   & 0 & 0 & 1 & 0 \\
 2 & 0 & 0 & 0 & 0 & 0 & 1 & 0 & 0 & 0 & 0 & 1 & 0 & 1 & 0 & 0 & 0 & 0 & 1 & 0
   & 0 & 0 & 0 & 0 \\
\hline
\end{tabular}
\caption{$\widehat{{N}}_{[2],g,\beta,J_r}(T_{2,3})$ for the trefoil}
 \end{minipage}
     \end{table}

 \begin{table}[h]\centering
   \resizebox{\columnwidth}{!}{
\begin{tabular}{|c|ccccc|ccccccc|ccccccccc|ccccccc|ccccc|}
\hline
 $\beta$& \multicolumn{5}{c|}{1} & \multicolumn{7}{c|}{2} & \multicolumn{9}{c|}{3} & \multicolumn{7}{c|}{4} & \multicolumn{5}{c|}{5}  \\
 \hline $g\backslash 2J_r$ & 0 & 1 & 2 & 3 & 4 & 1 & 2 & 3 & 4 & 5 & 6 & 7 & 2 & 3 & 4
   & 5 & 6 & 7 & 8 & 9 & 10 & 5 & 6 & 7 & 8 & 9 & 10 & 11 & 8 & 9 & 10 & 11 &
 12 \\
 \hline
 0 & 1 & 0 & 2 & 0 & 1 & 2 & 0 & 7 & 0 & 5 & 0 & 2 & 1 & 0 & 8 & 0 & 9 & 0 & 5
   & 0 & 1 & 3 & 0 & 7 & 0 & 4 & 0 & 2 & 2 & 0 & 1 & 0 & 1 \\
 1 & 0 & 2 & 0 & 2 & 0 & 0 & 5 & 0 & 10 & 0 & 5 & 0 & 0 & 3 & 0 & 14 & 0 & 12
   & 0 & 3 & 0 & 0 & 6 & 0 & 9 & 0 & 5 & 0 & 0 & 2 & 0 & 2 & 0 \\
 2 & 0 & 0 & 1 & 0 & 0 & 0 & 0 & 4 & 0 & 4 & 0 & 0 & 0 & 0 & 3 & 0 & 8 & 0 & 3
   & 0 & 0 & 0 & 0 & 4 & 0 & 4 & 0 & 0 & 0 & 0 & 1 & 0 & 0 \\
 3 & 0 & 0 & 0 & 0 & 0 & 0 & 0 & 0 & 1 & 0 & 0 & 0 & 0 & 0 & 0 & 1 & 0 & 1 & 0
   & 0 & 0 & 0 & 0 & 0 & 1 & 0 & 0 & 0 & 0 & 0 & 0 & 0 & 0 \\
\hline
\end{tabular}
 }
\caption{$\widehat{{N}}_{[1,1],g,\beta,J_r}(T_{2,3})$ for the trefoil}
 \end{table}
 \begin{table}[h]\centering
   \resizebox{\columnwidth}{!}{
\begin{tabular}{|c|ccc|ccccccc|ccccccccccc|ccccccccccccc|ccccccccccccc|ccccccccccc|ccccccccc|}
\hline
 $\beta$& \multicolumn{3}{c|}{1} & \multicolumn{7}{c|}{2} & \multicolumn{11}{c|}{3} & \multicolumn{13}{c|}{4} & \multicolumn{13}{c|}{5} & \multicolumn{11}{c|}{6} & \multicolumn{9}{c|}{7} \\
 \hline $g\backslash 2J_r$ & 0 & 1 & 2 & 1 & 2 & 3 & 4 & 5 & 6 & 7 & 2 & 3 & 4 & 5 & 6
   & 7 & 8 & 9 & 10 & 11 & 12 & 3 & 4 & 5 & 6 & 7 & 8 & 9 & 10 & 11 & 12 & 13
   & 14 & 15 & 6 & 7 & 8 & 9 & 10 & 11 & 12 & 13 & 14 & 15 & 16 & 17 & 18 & 9
   & 10 & 11 & 12 & 13 & 14 & 15 & 16 & 17 & 18 & 19 & 12 & 13 & 14 & 15 & 16
 & 17 & 18 & 19 & 20 \\
 \hline
 0 & 1 & 0 & 1 & 3 & 0 & 8 & 0 & 5 & 0 & 2 & 3 & 0 & 18 & 0 & 23 & 0 & 14 & 0
   & 5 & 0 & 1 & 1 & 0 & 16 & 0 & 38 & 0 & 34 & 0 & 19 & 0 & 6 & 0 & 2 & 5 & 0
   & 27 & 0 & 37 & 0 & 27 & 0 & 12 & 0 & 5 & 0 & 1 & 7 & 0 & 18 & 0 & 17 & 0 &
   10 & 0 & 4 & 0 & 2 & 3 & 0 & 4 & 0 & 3 & 0 & 1 & 0 & 1 \\
 1 & 0 & 1 & 0 & 0 & 6 & 0 & 10 & 0 & 5 & 0 & 0 & 9 & 0 & 37 & 0 & 38 & 0 & 18
   & 0 & 4 & 0 & 0 & 4 & 0 & 44 & 0 & 82 & 0 & 65 & 0 & 28 & 0 & 9 & 0 & 0 &
   17 & 0 & 69 & 0 & 83 & 0 & 53 & 0 & 24 & 0 & 5 & 0 & 0 & 20 & 0 & 43 & 0 &
   38 & 0 & 21 & 0 & 9 & 0 & 0 & 7 & 0 & 9 & 0 & 6 & 0 & 4 & 0 \\
 2 & 0 & 0 & 0 & 0 & 0 & 4 & 0 & 4 & 0 & 0 & 0 & 0 & 10 & 0 & 29 & 0 & 22 & 0
   & 6 & 0 & 0 & 0 & 0 & 6 & 0 & 48 & 0 & 72 & 0 & 45 & 0 & 16 & 0 & 0 & 0 & 0
   & 23 & 0 & 74 & 0 & 77 & 0 & 43 & 0 & 10 & 0 & 0 & 0 & 0 & 24 & 0 & 45 & 0
   & 36 & 0 & 16 & 0 & 0 & 0 & 0 & 7 & 0 & 9 & 0 & 6 & 0 & 0 \\
 3 & 0 & 0 & 0 & 0 & 0 & 0 & 1 & 0 & 0 & 0 & 0 & 0 & 0 & 5 & 0 & 10 & 0 & 4 &
   0 & 0 & 0 & 0 & 0 & 0 & 4 & 0 & 26 & 0 & 30 & 0 & 14 & 0 & 0 & 0 & 0 & 0 &
   0 & 16 & 0 & 42 & 0 & 35 & 0 & 10 & 0 & 0 & 0 & 0 & 0 & 0 & 16 & 0 & 25 & 0
   & 14 & 0 & 0 & 0 & 0 & 0 & 0 & 4 & 0 & 4 & 0 & 0 & 0 \\
\hline
\end{tabular}
 }
\caption{$\widehat{{N}}_{[3],g,\beta,J_r}(T_{2,3})$ for the trefoil}
 \end{table}

 \begin{table}[h]\centering
   \resizebox{\columnwidth}{!}{
\begin{tabular}{|c|ccccccc|ccccccccccc|ccccccccccccccc|ccccccccccccccccc|ccccccccccccccccc|ccccccccccccccc|ccccccccccccc|}
\hline
 $\beta$ & \multicolumn{7}{c|}{1} & \multicolumn{11}{c|}{2} & \multicolumn{15}{c|}{3} & \multicolumn{17}{c|}{4} & \multicolumn{17}{c|}{5} & \multicolumn{15}{c|}{6} & \multicolumn{13}{c|}{7}  \\
 \hline $g\backslash 2J_r$ & 0 & 1 & 2 & 3 & 4 & 5 & 6 & 1 & 2 & 3 & 4 & 5 & 6 & 7 & 8
   & 9 & 10 & 11 & 2 & 3 & 4 & 5 & 6 & 7 & 8 & 9 & 10 & 11 & 12 & 13 & 14 & 15
   & 16 & 3 & 4 & 5 & 6 & 7 & 8 & 9 & 10 & 11 & 12 & 13 & 14 & 15 & 16 & 17 &
   18 & 19 & 6 & 7 & 8 & 9 & 10 & 11 & 12 & 13 & 14 & 15 & 16 & 17 & 18 & 19 &
   20 & 21 & 22 & 9 & 10 & 11 & 12 & 13 & 14 & 15 & 16 & 17 & 18 & 19 & 20 &
   21 & 22 & 23 & 12 & 13 & 14 & 15 & 16 & 17 & 18 & 19 & 20 & 21 & 22 & 23 &
 24 \\
 \hline
 0 & 2 & 0 & 5 & 0 & 3 & 0 & 1 & 8 & 0 & 30 & 0 & 33 & 0 & 19 & 0 & 7 & 0 & 2
   & 10 & 0 & 61 & 0 & 104 & 0 & 85 & 0 & 46 & 0 & 19 & 0 & 6 & 0 & 1 & 4 & 0
   & 52 & 0 & 142 & 0 & 158 & 0 & 110 & 0 & 57 & 0 & 25 & 0 & 8 & 0 & 2 & 16 &
   0 & 89 & 0 & 141 & 0 & 122 & 0 & 76 & 0 & 39 & 0 & 17 & 0 & 6 & 0 & 1 & 21
   & 0 & 59 & 0 & 63 & 0 & 47 & 0 & 27 & 0 & 14 & 0 & 6 & 0 & 2 & 9 & 0 & 12 &
   0 & 11 & 0 & 7 & 0 & 4 & 0 & 2 & 0 & 1 \\
 1 & 0 & 5 & 0 & 7 & 0 & 3 & 0 & 0 & 26 & 0 & 71 & 0 & 64 & 0 & 31 & 0 & 9 & 0
   & 0 & 39 & 0 & 188 & 0 & 269 & 0 & 203 & 0 & 102 & 0 & 35 & 0 & 6 & 0 & 0 &
   18 & 0 & 190 & 0 & 434 & 0 & 454 & 0 & 308 & 0 & 155 & 0 & 55 & 0 & 13 & 0
   & 0 & 66 & 0 & 301 & 0 & 446 & 0 & 387 & 0 & 244 & 0 & 119 & 0 & 42 & 0 & 7
   & 0 & 0 & 75 & 0 & 190 & 0 & 214 & 0 & 163 & 0 & 97 & 0 & 44 & 0 & 13 & 0 &
   0 & 26 & 0 & 42 & 0 & 39 & 0 & 27 & 0 & 15 & 0 & 6 & 0 \\
\hline
\end{tabular}
 }
\caption{$\widehat{{N}}_{[2,1],g,\beta,J_r}(T_{2,3})$ for the trefoil}
 \end{table}

 \begin{table}[h]\centering
   \resizebox{\columnwidth}{!}{
\begin{tabular}{|c|ccccccccc|ccccccccccccc|ccccccccccccccccc|ccccccccccccccccccc|ccccccccccccccccccc|ccccccccccccccccc|ccccccccccccccc|}
\hline
 $\beta$ & \multicolumn{9}{c|}{1} & \multicolumn{13}{c|}{2} & \multicolumn{17}{c|}{3} & \multicolumn{19}{c|}{4} & \multicolumn{19}{c|}{5} & \multicolumn{17}{c|}{6} & \multicolumn{15}{c|}{7}  \\
 \hline $g\backslash 2J_r$ & 0 & 1 & 2 & 3 & 4 & 5 & 6 & 7 & 8 & 1 & 2 & 3 & 4 & 5 & 6
   & 7 & 8 & 9 & 10 & 11 & 12 & 13 & 2 & 3 & 4 & 5 & 6 & 7 & 8 & 9 & 10 & 11 &
   12 & 13 & 14 & 15 & 16 & 17 & 18 & 3 & 4 & 5 & 6 & 7 & 8 & 9 & 10 & 11 & 12
   & 13 & 14 & 15 & 16 & 17 & 18 & 19 & 20 & 21 & 6 & 7 & 8 & 9 & 10 & 11 & 12
   & 13 & 14 & 15 & 16 & 17 & 18 & 19 & 20 & 21 & 22 & 23 & 24 & 9 & 10 & 11 &
   12 & 13 & 14 & 15 & 16 & 17 & 18 & 19 & 20 & 21 & 22 & 23 & 24 & 25 & 12 &
 13 & 14 & 15 & 16 & 17 & 18 & 19 & 20 & 21 & 22 & 23 & 24 & 25 & 26 \\
 \hline
 0 & 1 & 0 & 4 & 0 & 4 & 0 & 2 & 0 & 1 & 5 & 0 & 23 & 0 & 34 & 0 & 25 & 0 & 14
   & 0 & 5 & 0 & 2 & 7 & 0 & 46 & 0 & 94 & 0 & 91 & 0 & 62 & 0 & 32 & 0 & 14 &
   0 & 5 & 0 & 1 & 3 & 0 & 39 & 0 & 117 & 0 & 148 & 0 & 120 & 0 & 76 & 0 & 38
   & 0 & 19 & 0 & 6 & 0 & 2 & 12 & 0 & 68 & 0 & 119 & 0 & 114 & 0 & 85 & 0 &
   49 & 0 & 27 & 0 & 12 & 0 & 5 & 0 & 1 & 15 & 0 & 45 & 0 & 52 & 0 & 45 & 0 &
   30 & 0 & 17 & 0 & 10 & 0 & 4 & 0 & 2 & 6 & 0 & 9 & 0 & 9 & 0 & 7 & 0 & 4 &
   0 & 3 & 0 & 1 & 0 & 1 \\
 1 & 0 & 4 & 0 & 10 & 0 & 8 & 0 & 4 & 0 & 0 & 22 & 0 & 78 & 0 & 96 & 0 & 68 &
   0 & 31 & 0 & 11 & 0 & 0 & 34 & 0 & 184 & 0 & 321 & 0 & 303 & 0 & 197 & 0 &
   98 & 0 & 36 & 0 & 7 & 0 & 0 & 16 & 0 & 173 & 0 & 449 & 0 & 552 & 0 & 452 &
   0 & 281 & 0 & 146 & 0 & 52 & 0 & 15 & 0 & 0 & 57 & 0 & 279 & 0 & 468 & 0 &
   474 & 0 & 353 & 0 & 218 & 0 & 107 & 0 & 42 & 0 & 8 & 0 & 0 & 63 & 0 & 177 &
   0 & 227 & 0 & 201 & 0 & 142 & 0 & 86 & 0 & 39 & 0 & 15 & 0 & 0 & 22 & 0 &
   39 & 0 & 42 & 0 & 34 & 0 & 24 & 0 & 12 & 0 & 7 & 0 \\
\hline
\end{tabular}
 }
\caption{$\widehat{{N}}_{[1,1,1],g,\beta,J_r}(T_{2,3})$ for the trefoil}
 \end{table}
 \begin{table}[h]\centering
\begin{tabular}{|c|ccccc|ccccc|ccc|}
\hline
 $\beta$& \multicolumn{5}{c|}{1} & \multicolumn{5}{c|}{2} & \multicolumn{3}{c|}{3} \\
 \hline $g\backslash 2J_r$ & -1 & 0 & 1 & 2 & 3 & 0 & 1 & 2 & 3 & 4 & 3 & 4 & 5 \\
 \hline
 0 & 1 & 0 & 1 & 0 & 1 & 1 & 0 & 2 & 0 & 2 & 1 & 0 & 1 \\
 1 & 0 & 2 & 0 & 2 & 0 & 0 & 2 & 0 & 3 & 0 & 0 & 1 & 0 \\
 2 & 0 & 0 & 1 & 0 & 0 & 0 & 0 & 1 & 0 & 0 & 0 & 0 & 0 \\
\hline
\end{tabular}
     \caption{$\widehat{{N}}_{[1],g,\beta,J_r}(T_{2,5})$}
 \end{table}

\clearpage
\subsection*{Torus links}

  \begin{table}[h]
 \begin{minipage}[c]{7.5cm}\centering

 \begin{tabular}{|c|c|c|}
\hline
 $\beta$ & \multicolumn{1}{c|}{-1} & \multicolumn{1}{c|}{0} \\
 \hline $g\backslash 2J_r$ & -2 & -1 \\
 \hline
 0 & 1 & 1 \\
\hline
 \end{tabular}
    \caption{$\widehat{{N}}_{[1],[1],g,\beta,J_r}(T_{2,2})$ for the Hopf link}
          \end{minipage}
   \begin{minipage}[c]{7.5cm}\centering

\begin{tabular}{|c|c|c|}
\hline
 $\beta$& \multicolumn{1}{c|}{-1} & \multicolumn{1}{c|}{0} \\
 \hline $g\backslash 2J_r$ & -3 & -2 \\
 \hline
 0 & 1 & 1 \\
\hline
\end{tabular}
\caption{$\widehat{{N}}_{[2],[1],g,\beta,J_r}(T_{2,2})$ for the Hopf link}
 \end{minipage}
\begin{minipage}[c]{7.5cm}\centering
\begin{tabular}{|c|ccc|ccc|c|}
\hline
 $\beta$& \multicolumn{3}{c|}{-2} & \multicolumn{3}{c|}{-1} & \multicolumn{1}{c|}{0} \\
 \hline $g\backslash 2J_r$ & -7 & -6 & -5 & -6 & -5 & -4 & -3 \\
\hline
 0 & 1 & 0 & 2 & 1 & 0 & 3 & 1 \\
 1 & 0 & 1 & 0 & 0 & 1 & 0 & 0 \\
\hline
\end{tabular}

\caption{$\widehat{{N}}_{[2],[2],g,\beta,J_r}(T_{2,2})$ for the Hopf link}
\end{minipage}
    \end{table}
 \begin{table}[h]
 \begin{minipage}[c]{12cm}\centering
\begin{tabular}{|c|c|c|}
\hline
 $\beta$& \multicolumn{1}{c|}{-2} & \multicolumn{1}{c|}{-1} \\
 \hline $g\backslash 2J_r$ & -5 & -4 \\
 \hline
 0 & 1 & 1 \\
\hline
\end{tabular}

\caption{$\widehat{{N}}_{[2],[1,1],g,\beta,J_r}(T_{2,2})$ for the Hopf link}
 \end{minipage}
    \begin{minipage}[c]{12cm}\centering

\begin{tabular}{|c|ccc|ccc|c|}
\hline
 $\beta$& \multicolumn{3}{c|}{-1} & \multicolumn{3}{c|}{0} & \multicolumn{1}{c|}{1}  \\
 \hline $g\backslash 2J_r$ & -4 & -3 & -2 & -3 & -2 & -1 & 0 \\
 \hline
 0 & 1 & 0 & 2 & 1 & 0 & 3 & 1 \\
 1 & 0 & 1 & 0 & 0 & 1 & 0 & 0 \\
\hline
\end{tabular}
\caption{$\widehat{{N}}_{[1],[1],g,\beta,J_r}(T_{2,4})$}

    \end{minipage}
   \end{table}
\begin{table}[h]
   \begin{minipage}[c]{12cm}\centering
\begin{tabular}{|c|ccccc|ccccc|ccc|}
\hline
 $\beta$& \multicolumn{5}{c|}{-1} & \multicolumn{5}{c|}{0} & \multicolumn{3}{c|}{1}  \\
 \hline $g\backslash 2J_r$ & -7 & -6 & -5 & -4 & -3 & -6 & -5 & -4 & -3 & -2 & -3 & -2
 & -1 \\
 \hline
 0 & 1 & 0 & 2 & 0 & 3 & 1 & 0 & 3 & 0 & 5 & 1 & 0 & 2 \\
 1 & 0 & 2 & 0 & 3 & 0 & 0 & 2 & 0 & 4 & 0 & 0 & 1 & 0 \\
 2 & 0 & 0 & 1 & 0 & 0 & 0 & 0 & 1 & 0 & 0 & 0 & 0 & 0 \\
\hline
\end{tabular}
\caption{$\widehat{{N}}_{[2],[1],g,\beta,J_r}(T_{2,4})$}

   \end{minipage}
 \begin{minipage}[c]{12cm}\centering
\begin{tabular}{|c|ccc|ccc|c|}
\hline
 $\beta$& \multicolumn{3}{c|}{-1} & \multicolumn{3}{c|}{0} & \multicolumn{1}{c|}{1}  \\
 \hline $g\backslash 2J_r$ & -5 & -4 & -3 & -4 & -3 & -2 & -1 \\
\hline
 0 & 1 & 0 & 1 & 1 & 0 & 2 & 1 \\
 1 & 0 & 1 & 0 & 0 & 1 & 0 & 0 \\
\hline
\end{tabular}
  \caption{$\widehat{{N}}_{[1,1],[1],g,\beta,J_r}(T_{2,4})$}
   \end{minipage}
  \end{table}
\end{landscape}

\clearpage

\begin{landscape}
\subsection*{Non-torus knots}

 \begin{table}[h]\centering
\begin{tabular}{|c|c|ccc|ccc|c|}
\hline
$\beta$ & \multicolumn{1}{c|}{-2} & \multicolumn{3}{c|}{-1} & \multicolumn{3}{c|}{0} & \multicolumn{1}{c|}{1}  \\
\hline
$g \backslash 2J_r$ & -3 & -2 & -1 & 0 & -1 & 0 & 1 & 2  \\
\hline
0 & 1 & 2 & 1 & 1 & 1 & 1 & 2 & 1 \\
1 & 0 & 0 & 1 & 0 & 0 & 1 & 0 & 0 \\
\hline
\end{tabular}
\caption{$\widehat{N}_{[1],g,\beta,J_r}(\mathbf{4_1})$ for the figure-eight knot}\label{table:fig8-uncolor}
\end{table}
 \begin{table}[h]\centering
   \resizebox{\columnwidth}{!}{
\begin{tabular}{|c|ccc|ccccc|ccccccc|ccccc|ccccc|ccc|c|}
\hline
 $\beta$   & \multicolumn{3}{c|}{-3} & \multicolumn{5}{c|}{-2} & \multicolumn{7}{c|}{-1} & \multicolumn{5}{c|}{0} & \multicolumn{5}{c|}{1} & \multicolumn{3}{c|}{2} & \multicolumn{1}{c|}{3}  \\
 \hline
 $g\backslash 2J_r$ & -8 & -7 & -6 & -7 & -6 & -5 & -4 & -3 & -6 & -5 & -4 & -3
   & -2 & -1 & 0 & -3 & -2 & -1 & 0 & 1 & 0 & 1 & 2 & 3 & 4 & 3 & 4 & 5 & 6
 \\
 \hline
 0 & 1 & 0 & 1 & 2 & 0 & 4 & 1 & 2 & 1 & 0 & 5 & 2 & 5 & 1 & 1 & 2 & 1 & 4 &
   2 & 3 & 1 & 1 & 3 & 0 & 1 & 1 & 0 & 2 & 1 \\
 1 & 0 & 1 & 0 & 0 & 3 & 0 & 3 & 0 & 0 & 2 & 0 & 6 & 1 & 2 & 0 & 0 & 3 & 1 &
   3 & 0 & 0 & 1 & 0 & 1 & 0 & 0 & 1 & 0 & 0 \\
 2 & 0 & 0 & 0 & 0 & 0 & 1 & 0 & 0 & 0 & 0 & 1 & 0 & 1 & 0 & 0 & 0 & 0 & 1 &
   0 & 0 & 0 & 0 & 0 & 0 & 0 & 0 & 0 & 0 & 0 \\
\hline
\end{tabular}
 }
\caption{$\widehat{{N}}_{[2],g,\beta,J_r}(\mathbf{4_1})$ for the figure-eight knot}
\end{table}

 \begin{table}[h]\centering
   \resizebox{\columnwidth}{!}{
\begin{tabular}{|c|c|ccc|ccccc|ccccc|ccccccc|ccccc|ccc|}
\hline
 $\beta$  & \multicolumn{1}{c|}{-3} & \multicolumn{3}{c|}{-2} & \multicolumn{5}{c|}{-1} & \multicolumn{5}{c|}{0} & \multicolumn{7}{c|}{1} & \multicolumn{5}{c|}{2} & \multicolumn{3}{c|}{3}  \\
 \hline
 $g\backslash 2J_r$ & -6 & -5 & -4 & -3 & -4 & -3 & -2 & -1 & 0 & -1 & 0 & 1 &
 2 & 3 & 0 & 1 & 2 & 3 & 4 & 5 & 6 & 3 & 4 & 5 & 6 & 7 & 6 & 7 & 8 \\
 \hline
 0 & 1 & 2 & 0 & 1 & 1 & 0 & 3 & 1 & 1 & 3 & 2 & 4 & 1 & 2 & 1 & 1 & 5 & 2 &
   5 & 0 & 1 & 2 & 1 & 4 & 0 & 2 & 1 & 0 & 1 \\
 1 & 0 & 0 & 1 & 0 & 0 & 1 & 0 & 1 & 0 & 0 & 3 & 1 & 3 & 0 & 0 & 2 & 1 & 6 &
   0 & 2 & 0 & 0 & 3 & 0 & 3 & 0 & 0 & 1 & 0 \\
 2 & 0 & 0 & 0 & 0 & 0 & 0 & 0 & 0 & 0 & 0 & 0 & 1 & 0 & 0 & 0 & 0 & 1 & 0 &
   1 & 0 & 0 & 0 & 0 & 1 & 0 & 0 & 0 & 0 & 0 \\
\hline
\end{tabular}
  }
  \caption{$\widehat{{N}}_{[1,1],g,\beta,J_r}(\mathbf{4_1})$ for the figure-eight knot}
\end{table}


 \begin{table}[h]\centering
  \resizebox{\columnwidth}{!}{
\begin{tabular}{|c|ccccccccc|ccccccccccc|ccccccccccccc|ccccccccccccc|ccccccccccc|ccccc
cccc|ccccccc|ccccccc|ccccc|ccc|}
\hline
 $ \beta $ & \multicolumn{9}{c|}{-4} & \multicolumn{11}{c|}{-3} & \multicolumn{13}{c|}{-2} & \multicolumn{13}{c|}{-1} & \multicolumn{11}{c|}{0} & \multicolumn{9}{c|}{1} & \multicolumn{7}{c|}{2} & \multicolumn{7}{c|}{3} & \multicolumn{5}{c|}{4} & \multicolumn{3}{c|}{5}  \\
 \hline
 $g\backslash 2J_r$ & -17 & -16 & -15 & -14 & -13 & -12 & -11 & -10 & -9 & -16
   & -15 & -14 & -13 & -12 & -11 & -10 & -9 & -8 & -7 & -6 & -15 & -14 & -13
   & -12 & -11 & -10 & -9 & -8 & -7 & -6 & -5 & -4 & -3 & -12 & -11 & -10 &
   -9 & -8 & -7 & -6 & -5 & -4 & -3 & -2 & -1 & 0 & -9 & -8 & -7 & -6 & -5 &
   -4 & -3 & -2 & -1 & 0 & 1 & -4 & -3 & -2 & -1 & 0 & 1 & 2 & 3 & 4 & 1 & 2
   & 3 & 4 & 5 & 6 & 7 & 4 & 5 & 6 & 7 & 8 & 9 & 10 & 7 & 8 & 9 & 10 & 11 &
   10 & 11 & 12 \\
\hline
 0 & 1 & 0 & 1 & 0 & 3 & 0 & 4 & 0 & 3 & 2 & 0 & 4 & 0 & 10 & 1 & 17 & 2 & 18
   & 3 & 7 & 1 & 0 & 5 & 0 & 12 & 2 & 27 & 6 & 37 & 11 & 28 & 5 & 5 & 2 & 0 &
   6 & 1 & 19 & 6 & 34 & 14 & 41 & 14 & 19 & 2 & 1 & 1 & 0 & 5 & 2 & 14 & 7 &
   26 & 13 & 26 & 6 & 6 & 2 & 1 & 6 & 4 & 15 & 6 & 11 & 1 & 2 & 3 & 2 & 8 & 2
   & 6 & 0 & 2 & 2 & 1 & 6 & 0 & 5 & 0 & 1 & 2 & 0 & 4 & 0 & 2 & 1 & 0 & 1 \\
 1 & 0 & 4 & 0 & 6 & 0 & 9 & 0 & 7 & 0 & 0 & 9 & 0 & 21 & 0 & 38 & 2 & 43 & 3
   & 20 & 0 & 0 & 5 & 0 & 24 & 0 & 53 & 5 & 83 & 12 & 69 & 9 & 17 & 0 & 0 & 9
   & 0 & 28 & 3 & 65 & 14 & 82 & 20 & 46 & 5 & 4 & 0 & 0 & 4 & 0 & 18 & 5 &
   38 & 13 & 41 & 9 & 13 & 0 & 0 & 5 & 2 & 12 & 4 & 13 & 1 & 3 & 0 & 0 & 4 &
   1 & 6 & 0 & 3 & 0 & 0 & 3 & 0 & 6 & 0 & 2 & 0 & 0 & 3 & 0 & 3 & 0 & 0 & 1
   & 0 \\
\hline
\end{tabular} }
\caption{$\widehat{{N}}_{[3],g,\beta,J_r}(\mathbf{4_1})$ for the figure-eight knot}
\end{table}


 \begin{table}[h]\centering
  \resizebox{\columnwidth}{!}{
\begin{tabular}{|c|ccccccc|ccccccccc|ccccccccccc|ccccccccccc|ccccccccccc|ccccccccccc|cc
 ccccccccc|ccccccccccc|ccccccccc|ccccccc|}
\hline
 $ \beta$  & \multicolumn{7}{c|}{-4} & \multicolumn{9}{c|}{-3} & \multicolumn{11}{c|}{-2} & \multicolumn{11}{c|}{-1} & \multicolumn{11}{c|}{0} & \multicolumn{11}{c|}{1} & \multicolumn{11}{c|}{2} & \multicolumn{11}{c|}{3} & \multicolumn{9}{c|}{4} & \multicolumn{7}{c|}{5} \\
 \hline
 $g\backslash 2J_r$ & -15 & -14 & -13 & -12 & -11 & -10 & -9 & -14 & -13 & -12
   & -11 & -10 & -9 & -8 & -7 & -6 & -13 & -12 & -11 & -10 & -9 & -8 & -7 &
   -6 & -5 & -4 & -3 & -10 & -9 & -8 & -7 & -6 & -5 & -4 & -3 & -2 & -1 & 0 &
   -7 & -6 & -5 & -4 & -3 & -2 & -1 & 0 & 1 & 2 & 3 & -2 & -1 & 0 & 1 & 2 & 3
   & 4 & 5 & 6 & 7 & 8 & 1 & 2 & 3 & 4 & 5 & 6 & 7 & 8 & 9 & 10 & 11 & 4 & 5
   & 6 & 7 & 8 & 9 & 10 & 11 & 12 & 13 & 14 & 7 & 8 & 9 & 10 & 11 & 12 & 13 &
 14 & 15 & 10 & 11 & 12 & 13 & 14 & 15 & 16 \\
 \hline
 0 & 1 & 0 & 2 & 0 & 4 & 0 & 4 & 2 & 0 & 6 & 0 & 14 & 1 & 19 & 2 & 8 & 1 & 0
   & 6 & 0 & 17 & 2 & 33 & 7 & 29 & 5 & 7 & 2 & 0 & 8 & 1 & 25 & 8 & 40 & 14
   & 28 & 5 & 6 & 1 & 0 & 7 & 3 & 25 & 13 & 41 & 15 & 26 & 4 & 6 & 6 & 4 & 26
   & 15 & 41 & 13 & 25 & 3 & 7 & 0 & 1 & 6 & 5 & 28 & 14 & 40 & 8 & 25 & 1 &
   8 & 0 & 2 & 7 & 5 & 29 & 7 & 33 & 2 & 17 & 0 & 6 & 0 & 1 & 8 & 2 & 19 & 1
   & 14 & 0 & 6 & 0 & 2 & 4 & 0 & 4 & 0 & 2 & 0 & 1 \\
 1 & 0 & 3 & 0 & 6 & 0 & 6 & 0 & 0 & 7 & 0 & 20 & 0 & 31 & 1 & 18 & 0 & 0 & 4
   & 0 & 21 & 0 & 47 & 3 & 53 & 5 & 17 & 0 & 0 & 7 & 0 & 25 & 2 & 52 & 10 &
   46 & 7 & 13 & 0 & 0 & 3 & 0 & 17 & 5 & 42 & 13 & 41 & 6 & 13 & 0 & 0 & 13
   & 6 & 41 & 13 & 42 & 5 & 17 & 0 & 3 & 0 & 0 & 13 & 7 & 46 & 10 & 52 & 2 &
   25 & 0 & 7 & 0 & 0 & 17 & 5 & 53 & 3 & 47 & 0 & 21 & 0 & 4 & 0 & 0 & 18 &
   1 & 31 & 0 & 20 & 0 & 7 & 0 & 0 & 6 & 0 & 6 & 0 & 3 & 0 \\
\hline
\end{tabular}   }
\caption{$\widehat{{N}}_{[2,1],g,\beta,J_r}(\mathbf{4_1})$ for the figure-eight knot}
 \end{table}


  \begin{table}[h]\centering
   \resizebox{\columnwidth}{!}{
\begin{tabular}{|c|ccc|ccccc|ccccccc|ccccccc|ccccccccc|ccccccccccc|ccccccccccccc|ccccccc
   cccccc|ccccccccccc|ccccccccc|}
\hline
$\beta$  & \multicolumn{3}{c|}{-4} & \multicolumn{5}{c|}{-3} & \multicolumn{7}{c|}{-2} & \multicolumn{7}{c|}{-1} & \multicolumn{9}{c|}{0} & \multicolumn{11}{c|}{1} & \multicolumn{13}{c|}{2} & \multicolumn{13}{c|}{3} & \multicolumn{11}{c|}{4} & \multicolumn{9}{c|}{5}  \\
 \hline
 $g\backslash 2J_r$ & -11 & -10 & -9 & -10 & -9 & -8 & -7 & -6 & -9 & -8 & -7 &
   -6 & -5 & -4 & -3 & -6 & -5 & -4 & -3 & -2 & -1 & 0 & -3 & -2 & -1 & 0 & 1
   & 2 & 3 & 4 & 5 & 0 & 1 & 2 & 3 & 4 & 5 & 6 & 7 & 8 & 9 & 10 & 1 & 2 & 3 &
   4 & 5 & 6 & 7 & 8 & 9 & 10 & 11 & 12 & 13 & 4 & 5 & 6 & 7 & 8 & 9 & 10 &
   11 & 12 & 13 & 14 & 15 & 16 & 7 & 8 & 9 & 10 & 11 & 12 & 13 & 14 & 15 & 16
 & 17 & 10 & 11 & 12 & 13 & 14 & 15 & 16 & 17 & 18 \\
 \hline
 0 & 1 & 0 & 1 & 2 & 0 & 4 & 0 & 2 & 1 & 0 & 5 & 0 & 6 & 1 & 2 & 2 & 0 & 6 &
   2 & 8 & 2 & 3 & 2 & 1 & 11 & 6 & 15 & 4 & 6 & 1 & 2 & 6 & 6 & 26 & 13 & 26
   & 7 & 14 & 2 & 5 & 0 & 1 & 1 & 2 & 19 & 14 & 41 & 14 & 34 & 6 & 19 & 1 & 6
   & 0 & 2 & 5 & 5 & 28 & 11 & 37 & 6 & 27 & 2 & 12 & 0 & 5 & 0 & 1 & 7 & 3 &
   18 & 2 & 17 & 1 & 10 & 0 & 4 & 0 & 2 & 3 & 0 & 4 & 0 & 3 & 0 & 1 & 0 & 1
   \\
 1 & 0 & 1 & 0 & 0 & 3 & 0 & 3 & 0 & 0 & 2 & 0 & 6 & 0 & 3 & 0 & 0 & 3 & 0 &
   6 & 1 & 4 & 0 & 0 & 3 & 1 & 13 & 4 & 12 & 2 & 5 & 0 & 0 & 13 & 9 & 41 & 13
   & 38 & 5 & 18 & 0 & 4 & 0 & 0 & 4 & 5 & 46 & 20 & 82 & 14 & 65 & 3 & 28 &
   0 & 9 & 0 & 0 & 17 & 9 & 69 & 12 & 83 & 5 & 53 & 0 & 24 & 0 & 5 & 0 & 0 &
   20 & 3 & 43 & 2 & 38 & 0 & 21 & 0 & 9 & 0 & 0 & 7 & 0 & 9 & 0 & 6 & 0 & 4
   & 0 \\
\hline
\end{tabular}   }
\caption{$\widehat{{N}}_{[1,1,1],g,\beta,J_r}(\mathbf{4_1})$ for the figure-eight knot}
\end{table}


 \begin{table}[h]\centering
\begin{tabular}{|c|ccc|cccc|ccc|c|}
\hline
 $ \beta$  & \multicolumn{3}{c|}{0} & \multicolumn{4}{c|}{1} & \multicolumn{3}{c|}{2} & \multicolumn{1}{c|}{3}\\
 \hline
 $g\backslash 2J_r$ & -1 & 0 & 1 & 0 & 1 & 2 & 3 & 2 & 3 & 4 & 5 \\
 \hline
 0 & 1 & 1 & 1 & 1 & 2 & 2 & 1 & 1 & 1 & 2 & 1 \\
 1 & 0 & 1 & 0 & 0 & 1 & 1 & 0 & 0 & 1 & 0 & 0 \\
\hline
\end{tabular}
\caption{$\widehat{{N}}_{[1],g,\beta,J_r}(\mathbf{5_2})$}
\end{table}

 \begin{table}[h]\centering
   \resizebox{\columnwidth}{!}{
\begin{tabular}{|c|ccc|cccccc|ccccccccc|cccccccccc|ccccccccccc|ccccccccc|ccccccc|}
\hline
 $\beta$  & \multicolumn{3}{c|}{1} & \multicolumn{6}{c|}{2} & \multicolumn{9}{c|}{3} & \multicolumn{10}{c|}{4} & \multicolumn{11}{c|}{5} & \multicolumn{9}{c|}{6} & \multicolumn{7}{c|}{7} \\
 \hline
 $g\backslash 2J_r$ & 0 & 1 & 2 & 1 & 2 & 3 & 4 & 5 & 6 & 2 & 3 & 4 & 5 & 6 & 7
   & 8 & 9 & 10 & 4 & 5 & 6 & 7 & 8 & 9 & 10 & 11 & 12 & 13 & 6 & 7 & 8 & 9 &
   10 & 11 & 12 & 13 & 14 & 15 & 16 & 9 & 10 & 11 & 12 & 13 & 14 & 15 & 16 &
 17 & 12 & 13 & 14 & 15 & 16 & 17 & 18 \\
 \hline
 0 & 1 & 1 & 1 & 2 & 4 & 5 & 4 & 3 & 1 & 1 & 5 & 9 & 13 & 12 & 7 & 4 & 2 & 1
   & 2 & 7 & 14 & 19 & 14 & 14 & 6 & 5 & 1 & 2 & 2 & 5 & 14 & 11 & 18 & 6 & 9
   & 2 & 5 & 0 & 1 & 4 & 3 & 10 & 2 & 7 & 1 & 4 & 0 & 2 & 2 & 0 & 2 & 0 & 1 &
   0 & 1 \\
 1 & 0 & 1 & 0 & 0 & 3 & 4 & 4 & 2 & 0 & 0 & 2 & 9 & 13 & 13 & 9 & 5 & 3 & 0
   & 0 & 5 & 15 & 20 & 28 & 14 & 17 & 3 & 7 & 0 & 0 & 6 & 9 & 29 & 12 & 29 &
   5 & 18 & 0 & 4 & 0 & 0 & 10 & 3 & 19 & 2 & 15 & 0 & 7 & 0 & 0 & 4 & 0 & 4
   & 0 & 3 & 0 \\
 2 & 0 & 0 & 0 & 0 & 0 & 1 & 1 & 0 & 0 & 0 & 0 & 1 & 5 & 5 & 4 & 3 & 0 & 0 &
   0 & 0 & 4 & 12 & 9 & 17 & 3 & 9 & 0 & 0 & 0 & 0 & 7 & 5 & 24 & 4 & 22 & 0
   & 6 & 0 & 0 & 0 & 0 & 10 & 1 & 16 & 0 & 9 & 0 & 0 & 0 & 0 & 3 & 0 & 3 & 0
   & 0 \\
 3 & 0 & 0 & 0 & 0 & 0 & 0 & 0 & 0 & 0 & 0 & 0 & 0 & 0 & 1 & 1 & 0 & 0 & 0 &
   0 & 0 & 0 & 1 & 5 & 1 & 5 & 0 & 0 & 0 & 0 & 0 & 0 & 4 & 1 & 10 & 0 & 4 & 0
   & 0 & 0 & 0 & 0 & 0 & 5 & 0 & 5 & 0 & 0 & 0 & 0 & 0 & 0 & 1 & 0 & 0 & 0 \\
 4 & 0 & 0 & 0 & 0 & 0 & 0 & 0 & 0 & 0 & 0 & 0 & 0 & 0 & 0 & 0 & 0 & 0 & 0 &
   0 & 0 & 0 & 0 & 0 & 1 & 0 & 0 & 0 & 0 & 0 & 0 & 0 & 0 & 1 & 0 & 1 & 0 & 0
   & 0 & 0 & 0 & 0 & 0 & 0 & 1 & 0 & 0 & 0 & 0 & 0 & 0 & 0 & 0 & 0 & 0 & 0 \\
\hline
\end{tabular}
 }
\caption{$\widehat{{N}}_{[2],g,\beta,J_r}(\mathbf{5_2})$}
\end{table}


 \begin{table}[h]\centering
   \resizebox{\columnwidth}{!}{
 \begin{tabular}{|c|ccccc|cccccccc|ccccccccccc|cccccccccccc|ccccccccccccc|
  ccccccccccc|
  ccccccccc|}
\hline
 $\beta$  & \multicolumn{5}{c|}{1} & \multicolumn{8}{c|}{2} & \multicolumn{11}{c|}{3} & \multicolumn{12}{c|}{4} & \multicolumn{13}{c|}{5} & \multicolumn{11}{c|}{6} & \multicolumn{9}{c|}{7}  \\
 \hline
 $g\backslash 2J_r$ & 0 & 1 & 2 & 3 & 4 & 1 & 2 & 3 & 4 & 5 & 6 & 7 & 8 & 2 & 3
   & 4 & 5 & 6 & 7 & 8 & 9 & 10 & 11 & 12 & 4 & 5 & 6 & 7 & 8 & 9 & 10 & 11 &
   12 & 13 & 14 & 15 & 6 & 7 & 8 & 9 & 10 & 11 & 12 & 13 & 14 & 15 & 16 & 17
   & 18 & 9 & 10 & 11 & 12 & 13 & 14 & 15 & 16 & 17 & 18 & 19 & 12 & 13 & 14
 & 15 & 16 & 17 & 18 & 19 & 20 \\
 \hline
 0 & 1 & 1 & 2 & 1 & 1 & 2 & 5 & 10 & 9 & 7 & 4 & 3 & 1 & 1 & 7 & 16 & 22 &
   21 & 17 & 13 & 7 & 4 & 2 & 1 & 3 & 10 & 21 & 30 & 26 & 24 & 14 & 14 & 6 &
   5 & 1 & 2 & 2 & 7 & 20 & 17 & 24 & 11 & 18 & 6 & 9 & 2 & 5 & 0 & 1 & 5 & 4
   & 13 & 3 & 10 & 2 & 7 & 1 & 4 & 0 & 2 & 3 & 0 & 2 & 0 & 2 & 0 & 1 & 0 & 1
   \\
 1 & 0 & 2 & 1 & 2 & 0 & 0 & 5 & 9 & 13 & 9 & 7 & 3 & 0 & 0 & 3 & 17 & 29 &
   35 & 32 & 22 & 14 & 7 & 4 & 0 & 0 & 9 & 26 & 42 & 58 & 41 & 49 & 21 & 24 &
   4 & 9 & 0 & 0 & 8 & 16 & 48 & 28 & 62 & 20 & 44 & 7 & 24 & 0 & 5 & 0 & 0 &
   15 & 6 & 33 & 6 & 32 & 3 & 21 & 0 & 9 & 0 & 0 & 6 & 0 & 8 & 0 & 6 & 0 & 4
   & 0 \\
 2 & 0 & 0 & 1 & 0 & 0 & 0 & 0 & 4 & 5 & 5 & 3 & 0 & 0 & 0 & 0 & 3 & 15 & 20
   & 21 & 16 & 9 & 6 & 0 & 0 & 0 & 0 & 10 & 28 & 32 & 53 & 25 & 39 & 6 & 16 &
   0 & 0 & 0 & 0 & 13 & 14 & 57 & 20 & 67 & 9 & 43 & 0 & 10 & 0 & 0 & 0 & 0 &
   20 & 4 & 41 & 3 & 36 & 0 & 16 & 0 & 0 & 0 & 0 & 7 & 0 & 9 & 0 & 6 & 0 & 0
   \\
 3 & 0 & 0 & 0 & 0 & 0 & 0 & 0 & 0 & 1 & 1 & 0 & 0 & 0 & 0 & 0 & 0 & 1 & 6 &
   6 & 5 & 4 & 0 & 0 & 0 & 0 & 0 & 0 & 5 & 17 & 11 & 26 & 4 & 14 & 0 & 0 & 0
   & 0 & 0 & 0 & 11 & 6 & 37 & 5 & 35 & 0 & 10 & 0 & 0 & 0 & 0 & 0 & 0 & 15 &
   1 & 25 & 0 & 14 & 0 & 0 & 0 & 0 & 0 & 0 & 4 & 0 & 4 & 0 & 0 & 0 \\
\hline
 \end{tabular}
 }
\caption{$\widehat{{N}}_{[1,1],g,\beta,J_r}(\mathbf{5_2})$}
\end{table}
 \begin{table}[h]\centering
\begin{tabular}{|c|c|ccc|cccc|ccc|c|}
\hline
 $\beta$  & \multicolumn{1}{c|}{-3} & \multicolumn{3}{c|}{-2} & \multicolumn{4}{c|}{-1} & \multicolumn{3}{c|}{0} & \multicolumn{1}{c|}{1} \\
 \hline
 $g\backslash 2J_r $& -5 & -4 & -3 & -2 & -3 & -2 & -1 & 0 & -1 & 0 & 1 & 2 \\
 \hline
 0 & 1 & 2 & 1 & 1 & 1 & 2 & 3 & 1 & 1 & 2 & 2 & 1 \\
 1 & 0 & 0 & 1 & 0 & 0 & 1 & 1 & 0 & 0 & 1 & 0 & 0 \\
\hline
\end{tabular}
\caption{$\widehat{{N}}_{[1],g,\beta,J_r}(\mathbf{6_1})$}
\end{table}


 \begin{table}[h]\centering
   \resizebox{\columnwidth}{!}{
\begin{tabular}{|c|ccccccc|ccccccccc|ccccccccccc|cccccccccc|ccccccccc|ccccccc|ccccc|ccc|c|}
\hline
 $\beta$ & \multicolumn{7}{c|}{-5} & \multicolumn{9}{c|}{-4} & \multicolumn{11}{c|}{-3} & \multicolumn{10}{c|}{-2} & \multicolumn{9}{c|}{-1} & \multicolumn{7}{c|}{0} & \multicolumn{5}{c|}{1} & \multicolumn{3}{c|}{2} & \multicolumn{1}{c|}{3}  \\
 \hline
 $g\backslash 2J_r$ & -16 & -15 & -14 & -13 & -12 & -11 & -10 & -15 & -14 & -13
   & -12 & -11 & -10 & -9 & -8 & -7 & -14 & -13 & -12 & -11 & -10 & -9 & -8 &
   -7 & -6 & -5 & -4 & -11 & -10 & -9 & -8 & -7 & -6 & -5 & -4 & -3 & -2 & -8
   & -7 & -6 & -5 & -4 & -3 & -2 & -1 & 0 & -4 & -3 & -2 & -1 & 0 & 1 & 2 & 0
 & 1 & 2 & 3 & 4 & 3 & 4 & 5 & 6 \\
 \hline
 0 & 1 & 0 & 1 & 0 & 2 & 0 & 2 & 2 & 0 & 4 & 1 & 7 & 2 & 10 & 3 & 4 & 1 & 0 &
   5 & 2 & 9 & 6 & 18 & 11 & 15 & 5 & 2 & 2 & 1 & 5 & 6 & 14 & 14 & 21 & 15 &
   8 & 2 & 1 & 2 & 4 & 7 & 13 & 15 & 11 & 7 & 1 & 1 & 3 & 5 & 6 & 8 & 3 & 1 &
   1 & 3 & 3 & 2 & 1 & 1 & 1 & 2 & 1 \\
 1 & 0 & 3 & 0 & 4 & 0 & 4 & 0 & 0 & 7 & 0 & 15 & 2 & 19 & 3 & 10 & 0 & 0 & 4
   & 0 & 18 & 5 & 29 & 12 & 29 & 9 & 6 & 0 & 0 & 7 & 3 & 17 & 14 & 28 & 20 &
   16 & 5 & 0 & 0 & 3 & 5 & 9 & 13 & 14 & 10 & 2 & 0 & 0 & 2 & 4 & 5 & 3 & 1
   & 0 & 0 & 1 & 1 & 1 & 0 & 0 & 1 & 0 & 0 \\
 2 & 0 & 0 & 3 & 0 & 3 & 0 & 0 & 0 & 0 & 9 & 0 & 16 & 1 & 10 & 0 & 0 & 0 & 0
   & 6 & 0 & 22 & 4 & 24 & 5 & 7 & 0 & 0 & 0 & 0 & 9 & 3 & 17 & 9 & 12 & 4 &
   0 & 0 & 0 & 0 & 3 & 4 & 5 & 5 & 1 & 0 & 0 & 0 & 0 & 1 & 1 & 0 & 0 & 0 & 0
   & 0 & 0 & 0 & 0 & 0 & 0 & 0 & 0 \\
 3 & 0 & 0 & 0 & 1 & 0 & 0 & 0 & 0 & 0 & 0 & 5 & 0 & 5 & 0 & 0 & 0 & 0 & 0 &
   0 & 4 & 0 & 10 & 1 & 4 & 0 & 0 & 0 & 0 & 0 & 0 & 5 & 1 & 5 & 1 & 0 & 0 & 0
   & 0 & 0 & 0 & 1 & 1 & 0 & 0 & 0 & 0 & 0 & 0 & 0 & 0 & 0 & 0 & 0 & 0 & 0 &
   0 & 0 & 0 & 0 & 0 & 0 & 0 \\
\hline
\end{tabular}
 }
\caption{$\widehat{{N}}_{[2],g,\beta,J_r}(\mathbf{6_1})$}
\end{table}


 \begin{table}[h]\centering
   \resizebox{\columnwidth}{!}{
\begin{tabular}{|c|ccccc|ccccccc|ccccccccc|cccccccc|cccccccc|ccccccc|ccccccc|ccccc|ccc|}
\hline
 $\beta$& \multicolumn{5}{c|}{-5} & \multicolumn{7}{c|}{-4} & \multicolumn{9}{c|}{-3} & \multicolumn{8}{c|}{-2} & \multicolumn{8}{c|}{-1} & \multicolumn{7}{c|}{0} & \multicolumn{7}{c|}{1} & \multicolumn{5}{c|}{2} & \multicolumn{3}{c|}{3}  \\
 \hline
 $g\backslash 2J_r$ & -14 & -13 & -12 & -11 & -10 & -13 & -12 & -11 & -10 & -9
   & -8 & -7 & -12 & -11 & -10 & -9 & -8 & -7 & -6 & -5 & -4 & -9 & -8 & -7 &
   -6 & -5 & -4 & -3 & -2 & -6 & -5 & -4 & -3 & -2 & -1 & 0 & 1 & -2 & -1 & 0
   & 1 & 2 & 3 & 4 & 0 & 1 & 2 & 3 & 4 & 5 & 6 & 3 & 4 & 5 & 6 & 7 & 6 & 7 &
 8 \\
 \hline
 0 & 1 & 0 & 1 & 0 & 2 & 2 & 0 & 4 & 1 & 7 & 2 & 3 & 1 & 0 & 5 & 2 & 9 & 6 &
   10 & 3 & 1 & 2 & 1 & 5 & 6 & 11 & 8 & 5 & 2 & 1 & 2 & 4 & 7 & 8 & 8 & 2 &
   2 & 2 & 5 & 10 & 6 & 7 & 2 & 1 & 1 & 4 & 6 & 8 & 5 & 2 & 1 & 2 & 3 & 4 & 1
   & 2 & 1 & 0 & 1 \\
 1 & 0 & 2 & 0 & 2 & 0 & 0 & 5 & 0 & 9 & 1 & 6 & 0 & 0 & 3 & 0 & 12 & 3 & 14
   & 4 & 3 & 0 & 0 & 5 & 2 & 10 & 7 & 7 & 3 & 0 & 0 & 2 & 3 & 4 & 6 & 2 & 3 &
   0 & 0 & 3 & 4 & 8 & 3 & 2 & 0 & 0 & 2 & 5 & 6 & 3 & 2 & 0 & 0 & 3 & 1 & 3
   & 0 & 0 & 1 & 0 \\
 2 & 0 & 0 & 1 & 0 & 0 & 0 & 0 & 4 & 0 & 4 & 0 & 0 & 0 & 0 & 3 & 0 & 8 & 1 &
   3 & 0 & 0 & 0 & 0 & 4 & 1 & 4 & 1 & 0 & 0 & 0 & 0 & 1 & 1 & 0 & 1 & 0 & 0
   & 0 & 0 & 1 & 1 & 1 & 0 & 0 & 0 & 0 & 1 & 1 & 1 & 0 & 0 & 0 & 0 & 1 & 0 &
   0 & 0 & 0 & 0 \\
 3 & 0 & 0 & 0 & 0 & 0 & 0 & 0 & 0 & 1 & 0 & 0 & 0 & 0 & 0 & 0 & 1 & 0 & 1 &
   0 & 0 & 0 & 0 & 0 & 0 & 1 & 0 & 0 & 0 & 0 & 0 & 0 & 0 & 0 & 0 & 0 & 0 & 0
   & 0 & 0 & 0 & 0 & 0 & 0 & 0 & 0 & 0 & 0 & 0 & 0 & 0 & 0 & 0 & 0 & 0 & 0 &
   0 & 0 & 0 & 0 \\
\hline
\end{tabular}
 }
\caption{$\widehat{{N}}_{[1,1],g,\beta,J_r}(\mathbf{6_1})$}
\end{table}


 \begin{table}[h]\centering
\begin{tabular}{|c|ccc|ccccc|ccccc|ccc|}
\hline
 $\beta$
 & \multicolumn{3}{c|}{-2} & \multicolumn{5}{c|}{-1} & \multicolumn{5}{c|}{0} & \multicolumn{3}{c|}{1}\\
 \hline
 $g\backslash 2J_r$ & -4 & -3 & -2 & -3 & -2 & -1 & 0 & 1 & -2 & -1 & 0 & 1 & 2
 & 1 & 2 & 3 \\
 \hline
 0 & 1 & 1 & 1 & 2 & 2 & 4 & 1 & 1 & 1 & 1 & 4 & 2 & 2 & 1 & 1 & 1 \\
 1 & 0 & 1 & 0 & 0 & 3 & 1 & 2 & 0 & 0 & 2 & 1 & 3 & 0 & 0 & 1 & 0 \\
 2 & 0 & 0 & 0 & 0 & 0 & 1 & 0 & 0 & 0 & 0 & 1 & 0 & 0 & 0 & 0 & 0 \\
\hline
\end{tabular}
\caption{$\widehat{{N}}_{[1],g,\beta,J_r}(\mathbf{6_3})$}
\end{table}


 \begin{table}[h]\centering
   \resizebox{\columnwidth}{!}{
\begin{tabular}{|c|ccccccccc|ccccccccccc|ccccccccccccc|ccccccccccccc|ccccccccccccc|ccc
   cccccc|ccccccc|}
\hline
 $\beta$ & \multicolumn{9}{c|}{-3} & \multicolumn{11}{c|}{-2} & \multicolumn{13}{c|}{-1} & \multicolumn{13}{c|}{0} & \multicolumn{13}{c|}{1} & \multicolumn{9}{c|}{2} & \multicolumn{7}{c|}{3}  \\
 \hline
 $g\backslash 2J_r$ & -12 & -11 & -10 & -9 & -8 & -7 & -6 & -5 & -4 & -11 & -10
   & -9 & -8 & -7 & -6 & -5 & -4 & -3 & -2 & -1 & -10 & -9 & -8 & -7 & -6 &
   -5 & -4 & -3 & -2 & -1 & 0 & 1 & 2 & -7 & -6 & -5 & -4 & -3 & -2 & -1 & 0
   & 1 & 2 & 3 & 4 & 5 & -4 & -3 & -2 & -1 & 0 & 1 & 2 & 3 & 4 & 5 & 6 & 7 &
 8 & 1 & 2 & 3 & 4 & 5 & 6 & 7 & 8 & 9 & 4 & 5 & 6 & 7 & 8 & 9 & 10 \\
 \hline
 0 & 1 & 1 & 1 & 2 & 4 & 4 & 4 & 3 & 2 & 2 & 2 & 4 & 6 & 12 & 13 & 19 & 16 &
   15 & 5 & 3 & 1 & 1 & 5 & 6 & 13 & 15 & 31 & 28 & 36 & 19 & 17 & 3 & 2 & 2
   & 2 & 6 & 7 & 21 & 20 & 38 & 27 & 33 & 11 & 9 & 2 & 2 & 1 & 1 & 5 & 5 & 18
   & 17 & 28 & 15 & 15 & 6 & 5 & 1 & 1 & 3 & 4 & 10 & 9 & 11 & 6 & 4 & 2 & 2
   & 1 & 2 & 3 & 2 & 1 & 1 & 1 \\
 1 & 0 & 4 & 3 & 6 & 6 & 10 & 6 & 6 & 0 & 0 & 9 & 7 & 21 & 20 & 40 & 29 & 40
   & 14 & 12 & 0 & 0 & 5 & 4 & 24 & 21 & 53 & 42 & 80 & 41 & 51 & 9 & 8 & 0 &
   0 & 9 & 7 & 27 & 22 & 63 & 39 & 74 & 25 & 31 & 5 & 7 & 0 & 0 & 4 & 3 & 17
   & 12 & 44 & 24 & 41 & 14 & 18 & 3 & 4 & 0 & 0 & 9 & 8 & 21 & 12 & 15 & 5 &
   7 & 0 & 0 & 3 & 3 & 4 & 2 & 3 & 0 \\
 2 & 0 & 0 & 6 & 3 & 9 & 4 & 7 & 0 & 0 & 0 & 0 & 16 & 9 & 36 & 20 & 44 & 14 &
   19 & 0 & 0 & 0 & 0 & 10 & 6 & 43 & 25 & 73 & 32 & 61 & 10 & 13 & 0 & 0 & 0
   & 0 & 16 & 9 & 42 & 21 & 62 & 18 & 37 & 4 & 9 & 0 & 0 & 0 & 0 & 6 & 3 & 20
   & 8 & 34 & 9 & 22 & 3 & 6 & 0 & 0 & 0 & 0 & 10 & 5 & 16 & 4 & 9 & 0 & 0 &
   0 & 0 & 3 & 1 & 3 & 0 & 0 \\
\hline
\end{tabular}
 }
\caption{$\widehat{{N}}_{[2],g,\beta,J_r}(\mathbf{6_3})$}
\end{table}

 \begin{table}[h]\centering
   \resizebox{\columnwidth}{!}{
\begin{tabular}{|c|ccccccc|ccccccccc|ccccccccccccc|ccccccccccccc|ccccccccccccc|cccccccc
   ccc|ccccccccc|}
\hline
 $\beta$& \multicolumn{7}{c|}{-3} & \multicolumn{9}{c|}{-2} & \multicolumn{13}{c|}{-1} & \multicolumn{13}{c|}{0} & \multicolumn{13}{c|}{1} & \multicolumn{11}{c|}{2} & \multicolumn{9}{c|}{3}  \\
 \hline
 $g\backslash 2J_r$ & -10 & -9 & -8 & -7 & -6 & -5 & -4 & -9 & -8 & -7 & -6 &
   -5 & -4 & -3 & -2 & -1 & -8 & -7 & -6 & -5 & -4 & -3 & -2 & -1 & 0 & 1 & 2
   & 3 & 4 & -5 & -4 & -3 & -2 & -1 & 0 & 1 & 2 & 3 & 4 & 5 & 6 & 7 & -2 & -1
   & 0 & 1 & 2 & 3 & 4 & 5 & 6 & 7 & 8 & 9 & 10 & 1 & 2 & 3 & 4 & 5 & 6 & 7 &
 8 & 9 & 10 & 11 & 4 & 5 & 6 & 7 & 8 & 9 & 10 & 11 & 12 \\
 \hline
 0 & 1 & 1 & 1 & 2 & 3 & 2 & 1 & 2 & 2 & 4 & 6 & 11 & 9 & 10 & 4 & 3 & 1 & 1
   & 5 & 6 & 15 & 15 & 28 & 17 & 18 & 5 & 5 & 1 & 1 & 2 & 2 & 9 & 11 & 33 &
   27 & 38 & 20 & 21 & 7 & 6 & 2 & 2 & 2 & 3 & 17 & 19 & 36 & 28 & 31 & 15 &
   13 & 6 & 5 & 1 & 1 & 3 & 5 & 15 & 16 & 19 & 13 & 12 & 6 & 4 & 2 & 2 & 2 &
   3 & 4 & 4 & 4 & 2 & 1 & 1 & 1 \\
 1 & 0 & 3 & 2 & 4 & 3 & 3 & 0 & 0 & 7 & 5 & 15 & 12 & 21 & 8 & 9 & 0 & 0 & 4
   & 3 & 18 & 14 & 41 & 24 & 44 & 12 & 17 & 3 & 4 & 0 & 0 & 7 & 5 & 31 & 25 &
   74 & 39 & 63 & 22 & 27 & 7 & 9 & 0 & 0 & 8 & 9 & 51 & 41 & 80 & 42 & 53 &
   21 & 24 & 4 & 5 & 0 & 0 & 12 & 14 & 40 & 29 & 40 & 20 & 21 & 7 & 9 & 0 & 0
   & 6 & 6 & 10 & 6 & 6 & 3 & 4 & 0 \\
 2 & 0 & 0 & 3 & 1 & 3 & 0 & 0 & 0 & 0 & 9 & 4 & 16 & 5 & 10 & 0 & 0 & 0 & 0
   & 6 & 3 & 22 & 9 & 34 & 8 & 20 & 3 & 6 & 0 & 0 & 0 & 0 & 9 & 4 & 37 & 18 &
   62 & 21 & 42 & 9 & 16 & 0 & 0 & 0 & 0 & 13 & 10 & 61 & 32 & 73 & 25 & 43 &
   6 & 10 & 0 & 0 & 0 & 0 & 19 & 14 & 44 & 20 & 36 & 9 & 16 & 0 & 0 & 0 & 0 &
   7 & 4 & 9 & 3 & 6 & 0 & 0 \\
\hline
\end{tabular}
 }
\caption{$\widehat{{N}}_{[1,1],g,\beta,J_r}(\mathbf{6_3})$}
\end{table}

\end{landscape}
\clearpage

\bibliography{CS}{}
\bibliographystyle{halpha}

\end{document}